\documentclass[aps,prd,10pt,superscriptaddress,twocolumn,nofootinbib,dvipsnames,longbibliography]{revtex4-2}

\usepackage[pdftex]{color}
\usepackage[sort&compress]{natbib}
\usepackage[colorlinks=true,linkcolor=purple,filecolor=orange,urlcolor=blue,citecolor=purple,pdftex,plainpages=false]{hyperref}
\usepackage{url}
\pdfoutput=1
\usepackage{ifpdf}
\usepackage[utf8]{inputenc}
\usepackage{color,hhline,psfrag,rotating}
\usepackage{dcolumn}
\usepackage{verbatim}
\usepackage{lipsum}
\usepackage{bm}
\usepackage{slashed}
\usepackage{braket}
\usepackage{rotating}
\usepackage{multirow,tabularx}
\usepackage{graphicx}
\usepackage[dvipsnames]{xcolor}
\usepackage{amsfonts,amssymb,amsmath}
\usepackage{booktabs}  

\mathchardef\mhyphen="2D

\begin{document}

\global\let\newpage\relax

\title{A practical guide to fitting correlation functions from lattice data}
\author{W.~G.~Parrott}
\email[]{wp251@cantab.ac.uk}
\affiliation{Department of Physics and Astronomy, York University, Toronto, Ontario, M3J 1P3, Canada}

\date{\today}
\begin{abstract}
This guide contains a collection of the tips, tricks, and techniques that we have found to be useful when performing very large, correlated Bayesian fits of two, and three-point correlation functions for semileptonic decays, in this case $B\to K$ and $D\to K$. It is written explicitly with the use of \textit{gvar}, \textit{lsqfit} and \textit{corrfitter} in mind, but many techniques are transferable to other fitting software, and other correlator fits. The guide is by no means comprehensive, and many problems can be approached from different angles and points of view. The intention here is simply to present ideas which may be useful to others.  
\end{abstract}

\maketitle
\section{Introduction}
In lattice QCD, we often find ourselves fitting two, and three-point correlation functions to extract amplitudes, energies, and matrix elements. As we are usually interested in pushing the limits of finer and more expensive lattices, it is frequently the case that we are working with a small fraction of the statistics that we require for an ideal fit. This makes the process of performing large, correlated fits a balancing act, with many choices and compromises to be made, typically between speed and uncertainty on posterior values.
\section{Data and fit forms}\label{sec:corrfits}
In our examples, we will focus on fits to $H\to K$ decay data, where $H=\bar{h}l$ for a heavy quark $h$ of mass $am_c\leq am_h\leq am_b$ (where $a$ is the lattice spacing) and light quark $l$ of mass $m_l$. The data shown in figures will come from a variety of Highly Improved Staggered Quark (HISQ)~\cite{Follana:2006rc} action ensembles, generated by the MILC collaboration~\cite{MILC:2010pul,MILC:2012znn}. These data sets have varying lattice spacings, $am_h$ values, $K$ momentum values (determined from twist $\theta$~\cite{Sachrajda:2004mi,Guadagnoli:2005be}), and lattice volumes $N_x^3\times N_t$. They contain different spin taste combinations of mesons (referred to as Goldstone and non-Goldstone), and scalar (S), temporal vector (V), spatial vector (X), and tensor (T) current insertions.  

Almost all of this information is irrelevant to the points we will be discussing, which generalise to any sets of similar lattice data, with the caveat that staggered quarks give rise to temporally oscillating terms in correlation functions~\cite{Follana:2006rc}, which are not present in other formalisms. For this reason, we will provide details about the specific nature of the data discussed only when it is relevant to the point being made. However, full details of the data used can be found in~\cite{Parrott:2022rgu}.

We use a standard Bayesian fitting approach~\cite{Lepage:2001ym}, and perform simultaneous, multi-exponential fits to both two, and three-point data using the Python packages \textit{gvar}, \textit{lsqfit} and \textit{corrfitter}~\cite{lsqfit,gvar,corrfitter}. These packages have extensive documentation which gives many examples of how to construct the code for fits explicitly, and we will not repeat this here.  

We fit two-point correlators for meson $M$ to a sum of exponentials representing a tower of possible states of energy $E^{M}_i$ and amplitude $d_i^{M}$,
\begin{equation}\label{eq:2ptcorr}
  \begin{split}
    C_2^M(t)&=\sum^{N_\mathrm{exp}}_{i=0}\big(|d_i^{M,\mathrm{n}}|^2(e^{-E_i^{M,\mathrm{n}}t}+e^{-E_i^{M,\mathrm{n}}(N_t-t)})\\
    &-(-1)^{t}|d_i^{M,\mathrm{o}}|^2(e^{-E_i^{M,\mathrm{o}}t}+e^{-E_i^{M,\mathrm{o}}(N_t-t)})\big).
  \end{split}
\end{equation}
The ground state is specified by $i=0$. Because of the nature of staggered quarks, we must also fit states which oscillate in time (labelled `o' as opposed to `n' for non-oscillating states).

We perform three point fits (for $H$ and $K$ mesons $M_2$ and $M_1$) with scalar, vector or tensor current insertions to the following form, 
\begin{equation}\label{eq:3ptcorr}
  \begin{split}
    &C^{M_1,M_2}_3(t,T)=\sum^{N_\mathrm{exp}}_{i,j=0}\big(d_i^{M_1,\mathrm{n}}J_{ij}^{\mathrm{nn}}d_j^{M_2,\mathrm{n}}e^{-E_i^{M_1,\mathrm{n}}t}e^{-E_j^{M_2,\mathrm{n}}(T-t)}\\
    &-(-1)^{(T-t)}d_i^{M_1,\mathrm{n}}J_{ij}^{\mathrm{no}}d_j^{M_2,\mathrm{o}}e^{-E_i^{M_1,\mathrm{n}}t}e^{-E_j^{M_2,\mathrm{o}}(T-t)}\\
    &-(-1)^{t}d_i^{M_1,\mathrm{o}}J_{ij}^{\mathrm{on}}d_j^{M_2,\mathrm{n}}e^{-E_i^{M_1,\mathrm{o}}t}e^{-E_j^{M_2,\mathrm{n}}(T-t)}\\
    &+(-1)^{T}d_i^{M_1,\mathrm{o}}J_{ij}^{\mathrm{oo}}d_j^{M_2,\mathrm{o}}e^{-E_i^{M_1,\mathrm{o}}t}e^{-E_j^{M_2,\mathrm{o}}(T-t)}\big).
  \end{split}
\end{equation}
Here $J_{ij}^{kl}$ ($i,j\in\{0,1,...,N_{\text{exp}}-1\}$, and $k,l\in\{\mathrm{n},\mathrm{o}\}$) are matrix elements of $J=S(V)[T]\{X\}$ for the scalar (temporal vector) [tensor] \{spatial vector\} currents. For example, $X_{ij}^{\mathrm{no}}$, gives the matrix element for the spatial vector current insertion ($X$) between the $i$th non-oscillating (`n') state of $M_1$ and the $j$th oscillating (`o') state of $M_2$.
$T$ and $t$ are the total source-sink time separation and the time of the current insertion, both in lattice units, where we have taken $t_0=0$ for simplicity. $T$ (with no sub/super script) is not to be confused with the tensor current insertion $T_{ij}^{kl}$.

Note that the $H$ meson here is $M_2$, at the sink, with the $K$ at the source. This time reversed setup makes no difference to the physics but makes computation easier.  

Both Eq.~\eqref{eq:2ptcorr} and Eq.~\eqref{eq:3ptcorr} are built in functions in \textit{corrfitter}~\cite{corrfitter}. These functions are then passed to \textit{lsqfit}~\cite{lsqfit}, which performs the fit. All correlations are tracked automatically using \textit{gvar}. See \textit{corrfitter} and \textit{lsqfit} documentation for more details.
\subsection{Priors}
In the Bayesian method we employ, all fit parameters in Eq.~\eqref{eq:2ptcorr} and Eq.~\eqref{eq:3ptcorr} require a prior. This permits fits to functions with more parameters than data points, as the fit parameters each have an associated prior, which acts somewhat like a data point, meaning the overall degrees of freedom (d.o.f.) in the fit is always equal to the number of data points, irrespective of the number of fit parameters. 
The fitter minimises $\chi^2=\chi^2_{\mathrm{data}}+\chi^2_{\mathrm{prior}}$~\cite{Lepage:2001ym}, which takes into account both the difference of the fitted points from the raw data, as well as the fitted parameters from their prior values. For this reason, prior selection is fundamental to the fitting process, and choosing reasonable priors is most of the challenge of fitting.
\subsection{SVD cuts}\label{sec:svd}
A full description of the need for SVD cuts is given in the \textit{corrfitter} documentation and Appendix D of~\cite{Dowdall:2019bea}, but the salient points are repeated here.

When the number of points in the data set $N_G$\footnote{For example, in a simple fit to one correlator of length $t$, $N_G=t$.} is not many times fewer than the number of random samples $N_s$ used to determine the data's covariance matrix\footnote{Typically, this would be the number of gauge configurations which are averaged over to make the data set.}, the estimation of the covariance matrix using the spread of samples can result in underestimation of the Eigenvalues, which can be thought of as an artificial reduction of uncertainty. In the extreme limit, where $N_G<N_s$, some Eigenvalues will necessarily be 0.
\begin{figure}
  \includegraphics[width=0.48\textwidth]{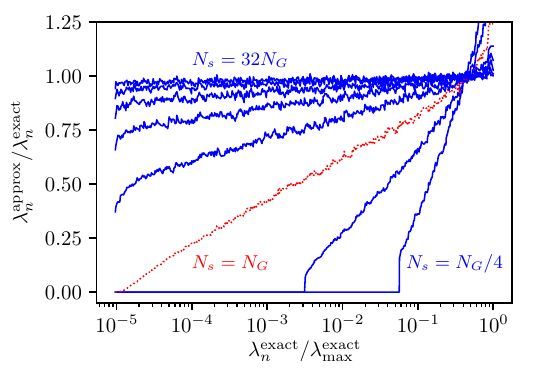}
  \caption{Taken from~\cite{Dowdall:2019bea} (Fig. 15), the variation in covariance matrix Eigenvalues calculated with different sample sizes $N_s$, relative to the known exact values.}
  \label{fig:Ns_dep}
\end{figure}

This is wonderfully illustrated in Fig.~15 of~\cite{Dowdall:2019bea}, which we reproduce here (Fig.~\ref{fig:Ns_dep}). This figure shows some example data for which the exact covariance matrix Eigenvalues $\lambda$ are known, and compares these with those calculated for different numbers of samples. For $N_s<32N_G$, some Eigenvalues are underestimated. In many real life cases, we work with $N_s\approx N_G$ or less, so this is a clear problem. The solution is an SVD cut. The data is bootstrapped and estimates for the ratio of the calculated Eigenvalues to the exact ones are made. An SVD cut is then applied, at the point where the estimates fall below $\lambda^{\mathrm{approx}}/\lambda^{\mathrm{exact}} <1-\sqrt{2/N_G}$. All Eigenvalues which are smaller than this are increased so that they equal the size of the Eigenvalue where the SVD cut is placed. This in necessarily a conservative move,  explicitly increasing uncertainty in the covariance matrix. The process of calculating and applying this SVD cut is handled by \textit{gvar}.  
    
\subsection{Noise}
Both the use of priors and the introduction of an SVD cut artificially reduce the $\chi^2$ of a fit. For an example in the case of priors, imagine choosing very broad priors for a fit parameter which is well determined by the data. The resulting $\chi^2_{\mathrm{prior}}$ contribution may be tiny, giving a reduced $\chi^2$.

Let's take a very simple, concrete example. Say we fit three data points to a straight line $Ax+B$. In the frequentist model, the number of degrees of freedom is $3~\mathrm{data~points} - 2~\mathrm{parameters} = 1$. Supposing the three data points lie on average $1\sigma$ from the best fit line, we then have $\chi^2=3$ and $\chi^2/\mathrm{d.o.f.}=3/1$ ($Q=0.08$), an acceptable, but not fantastic, fit.

Suppose we instead take a Bayesian approach with prior values for $A$ and $B$, and choose them so that they are extremely broad. We still have $\chi^2_{\mathrm{data}}=3$, but now we add on $\chi^2_{\mathrm{prior}}\approx 0$. Crucially, however, the priors alter the degrees of freedom, cancelling out the $-2$ fit parameters, so that $\mathrm{d.o.f.}=3$. This means our $\chi^2/\mathrm{d.o.f.}\approx 3/3$ ($Q=0.39$), so we have lowered $\chi^2$ (and raised the $Q$ value) artificially. 

The solution to this is to add prior and SVD noise, which are discussed in Appendix D of~\cite{Dowdall:2019bea}. The SVD noise modifies the data, choosing a random sample drawn from the distribution of the SVD contribution to the uncertainty in the data for every data point.
The prior noise is similar, modifying prior central values to a random value from with the prior's specified distribution. To continue the example above, on average, the random variation of the prior would lead to $\chi^2_{\mathrm{prior}}= 2$ with the noise on. This then gives a larger $\chi^2/\mathrm{d.o.f.}=5/3$ ($Q=0.17$). The $\chi^2$ is still smaller than in the frequentist approach, reflecting the fact that we have some prior knowledge and our fit agrees with this. If instead we had tried to fit the data to a quadratic, we'd have an additional prior value, and $\chi^2/\mathrm{d.o.f.}=6/3$ ($Q=0.11$). If we add 10 redundant parameters, we get $\chi^2/\mathrm{d.o.f.}=15/3$ ($Q=0.002$).

If the SVD cut and priors are reasonable, a good fit should return a $\chi^2/\mathrm{d.o.f.}$ close to 1 ($Q\gtrsim 0.05$) with this noise turned on, as well as having fit results that are consistent with the noiseless ones.  

As shown above, fits with large numbers of parameters relative to data points have their $\chi^2/\mathrm{d.o.f.}$ increased because the degrees of freedom does not include parameters which have priors. This penalises over-fitting. The Gaussian Bayes factor (GBF) is also computed by \textit{lsqfit}, and tells us about the likelihood that the data was derived from the fit form. This also penalises over-fitting.

\subsection{Objectives}
Given everything discussed so far, the objective is to make choices for priors and other fit parameters which yield a fit which is stable against changes in the choices we make, maximises the GBF and gives a good $\chi^2$ when noise is applied. As secondary objectives, a fit which runs in finite time and minimises uncertainty on posterior values is also attractive.

The fitting process is all about compromise. Reducing the size of the data set by limiting the $t$ range of fitted data, and other techniques below, reduces the size of the covariance matrix, and so the SVD cut, potentially reducing uncertainty. It also speeds up the fit. On the other hand, a reduction in data points for an unchanged number of priors will push up $\chi^2/\mathrm{d.o.f.}$, all other things being equal, and can increase uncertainty if we are throwing away useful data.

\section{Prior selection}
As stated above, choosing priors which are reasonable is the bulk of performing a fit. Some priors can be quite precisely estimated, but others are more order of magnitude guesses. It's essential that we test our prior choices to check that they are not having an outsized effect on the final results, particularly if they are priors we don't have a good handle on in the first place.
A key first step in choosing priors is to accrue all possible prior information. We can then use this to make educated guesses for as many parameters as possible. For our case, let's write down some things we know:
\begin{itemize}
  \item We can use effective mass ($E_0^{\mathrm{n,eff}}$) and amplitude ($d_0^{\mathrm{n,eff}}$) plots to estimate the ground state energies and amplitudes of all two-point correlators
  \item We can use these effective masses and amplitudes to calculate an effective three-point amplitude $J_{00}^{\mathrm{nn,eff}}$
  \item Excited state energy splittings are $\mathcal{O}(\Lambda_{\mathrm{QCD}})$
  \item Oscillating-non-oscillating ground state energy splittings are $\mathcal{O}(\Lambda_{\mathrm{QCD}})$
  \item Amplitudes for excited and oscillating states are likely to be less than the ground state, or at most the same order of magnitude
  \item $J_{ij\neq 00}^{kl}$ and $J_{00}^{kl\neq \mathrm{nn}}$ are likely to be less than $J_{00}^{\mathrm{nn}}$, or at most the same order of magnitude
  \item Looking at log plots of the two-point correlators, we can estimate the relative size of oscillating state amplitudes
  \item Energies and amplitudes from mesons (in this case $K$) with momenta can be related via the relativistic dispersion relation
\end{itemize}
\subsection{Reducing the number of variable priors}
As an additional step, we want to ultimately be able to perform a stability analysis where we vary all our priors and check the effects on the fit - it's much more practical to do this if we are able to link groups of priors together in sensible ways, so we can adjust a large number of priors using just one parameter.

For example, rather than estimating priors for all excited state and oscillating amplitudes, we can link them to our effective ground state amplitude\footnote{A full description of calculating ground state effective masses and amplitudes will be given in the example below.}, $P[d^{\mathrm{n/o}}_{i\neq0}]=Ad^{\mathrm{n},\mathrm{eff}}_0 \pm Bd^{\mathrm{n},\mathrm{eff}}_0$, as $d$ is always positive (in our case)\footnote{For a $d$ which can be negative, we could do the same this but with a central value of 0.}. Now, rather than having to vary $d_{i\neq0}$ for every two-point correlator in the fit, we just have two parameters to play with, $A$ and $B$. Based on our assumptions above $A=0.5$ and $B=0.5$ might be a good choice, and variations around that can be trialled. Given that we know little about these priors, except that the excited states are likely to be of the same order of magnitude as the ground state, and probably smaller, it's not unreasonable to join all these priors together, as we have no specific prior knowledge about any of them being larger or smaller than any others, beyond the size of the ground state effective amplitude, which is automatically accounted for.

It is of course crucial that these priors are not `the same' in the sense of being the same \textbf{gvar.GVar} object\footnote{See \textit{gvar} documentation if this is unfamiliar notation.} and so 100\% correlated, but rather distinct \textbf{gvar.GVar}s with the same value. 

We can play a similar game with the priors for $J_{00}^{kl\neq \mathrm{nn}}$ and $J_{ij\neq 00}^{kl}$, but this time the values can be positive or negative. The former we can assume are of the order of magnitude of $J_{00}^{\mathrm{nn,eff}}$, giving $P[J_{00}^{kl\neq \mathrm{nn}}] = 0\pm CJ_{00}^{\mathrm{nn,eff}}$, where $C=1$ is a good starting point. Excited state overlaps we may expect to be somewhat smaller, say $P[J_{ij\neq 00}^{kl}] = 0\pm DJ_{00}^{\mathrm{nn,eff}}$ with $D=0.5$. Again, we have reduced the number of parameters to trial down to just two, covering a huge number of priors.
\subsection{Building in the relativistic dispersion relation}
If our fit contains multiple mesons of different momenta, we might also choose to build the relativistic dispersion relation into our fits. This is a way of relating the ground state amplitude and energy priors for mesons with finite momentum to those with zero momentum. It may not be that useful in cases where we have good estimates for the effective masses and amplitudes anyway, but high momentum meson correlators are often noisy, and effective masses and amplitudes may be less precise that priors based off the dispersion relation. For the energies, the relation in terms of momentum $\vec{p}$ is simply the relativistic dispersion relation,
\begin{equation}\label{Eq:dispE}
  P[aE_{i,\vec{p}}]=\sqrt{P[aE_{i,\vec{0}}]^2+(a\vec{p})^2}\Big(1+P[c_2]\Big(\frac{a\vec{p}}{\pi}\Big)^2\Big),
\end{equation}
where we set prior $P[c_2]=0(1)$ to allow for momentum dependent discretisation effects, and we may choose to add (typically negligible) $(a\vec{p}/\pi)^4$ terms.

For the amplitude, the derivation is more complex. We start by defining the Lorentz invariant overlap factor (taking the $K$ as our example) $\langle \Phi^K|K_{i,\vec{p}}\rangle$, which is defined by
\begin{equation}
  \langle \Phi^K|K_{i,\vec{p}}\rangle = \frac{1}{a^2}d^{K}_{i,\vec{p}}\sqrt{aE^K_{i,\vec{p}}}.
\end{equation}
Setting these factors equal for momenta $\vec{0}$ and $\vec{p}$ yields,
\begin{equation}
  \begin{split}
    d^{K}_{i,\vec{0}}\sqrt{aE^K_{i,\vec{0}}}&=d^{K}_{i,\vec{p}}\sqrt{aE^K_{i,\vec{p}}}\\
    &= d^{K}_{i,\vec{p}}\big((aE^K_{i,\vec{0}})^2+(a\vec{p})^2\big)^{1/4}\\
    \implies d^{K}_{i,\vec{0}}&= d^{K}_{i,\vec{p}}\big(1+(a\vec{p}/aE^K_{i,\vec{0}})^2\big)^{1/4}.
  \end{split}
\end{equation}
Using this expression we can again relate priors of different momenta,
\begin{equation}\label{Eq:dispa}
  P[d_{0,\vec{p}}]=\frac{P[d_{0,\vec{0}}]}{[1+(a\vec{p}/P[aE_{0,\vec{0}}])^2]^{1/4}}\Big(1+P[d_2]\Big(\frac{a\vec{p}}{\pi}\Big)^2\Big),
\end{equation}
where, we again add in discretisation effects. If we follow the derivation more carefully, we find that (neglecting higher order terms) $\Big(1+P[d_2]\Big(\frac{a\vec{p}}{\pi}\Big)^2\Big)$ should be equal to $\Big(1+P[c_2]\Big(\frac{a\vec{p}}{\pi}\Big)^2\Big)^{-1/2}$, giving $d_2=-c_2/2$, which we may or may not choose to enforce in our fits. 
These relations can be applied to oscillating and non-oscillating priors alike. 
\subsection{Non-Gaussian priors}
If source and sink operators are the same, as in our case, then without loss of generality, we can impose $d^{\mathrm{n/o}}_i>0$. This doesn't change any results (up to a sign), but vastly speeds up the fit, as it doesn't have to trial negative values which would give the same $\chi^2$. Similarly, all $E_i$ values are positive. In fact, \textit{corrfitter} does not take absolute values for $E_{i\neq0}$, but rather splittings, which the fitter could potentially find to be negative if positivity is not enforced.

A typical solution to this is to use logarithmic priors for positive definite values. Rather than fitting a Gaussian distribution for $d_i$, for example, \textit{corrfitter} is presented with a Gaussian distribution for $\log(d_i)$, which is exponentiated in the fit to give $d_i$, implying a logarithmic distribution for the amplitude, which may not be in line with our expectations. This typically works smoothly in the fit, but it can lead to issues with noise. Another alternative is to use a square root distribution, where$\sqrt{d_i}$ is Gaussian distributed. 
\begin{figure}
  \includegraphics[width=0.44\textwidth]{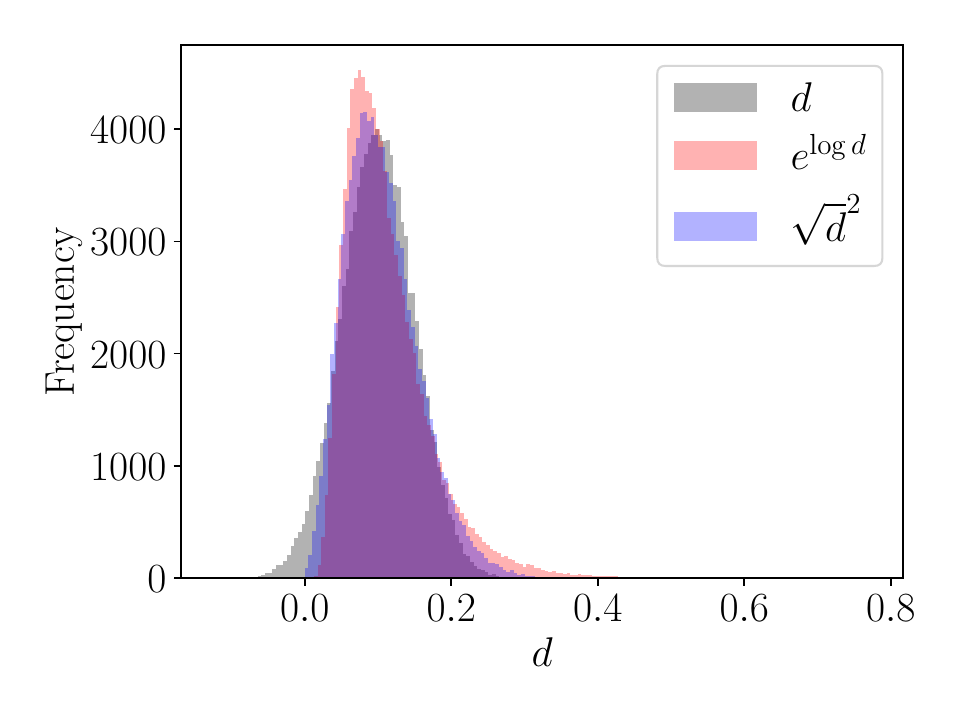}
  \includegraphics[width=0.44\textwidth]{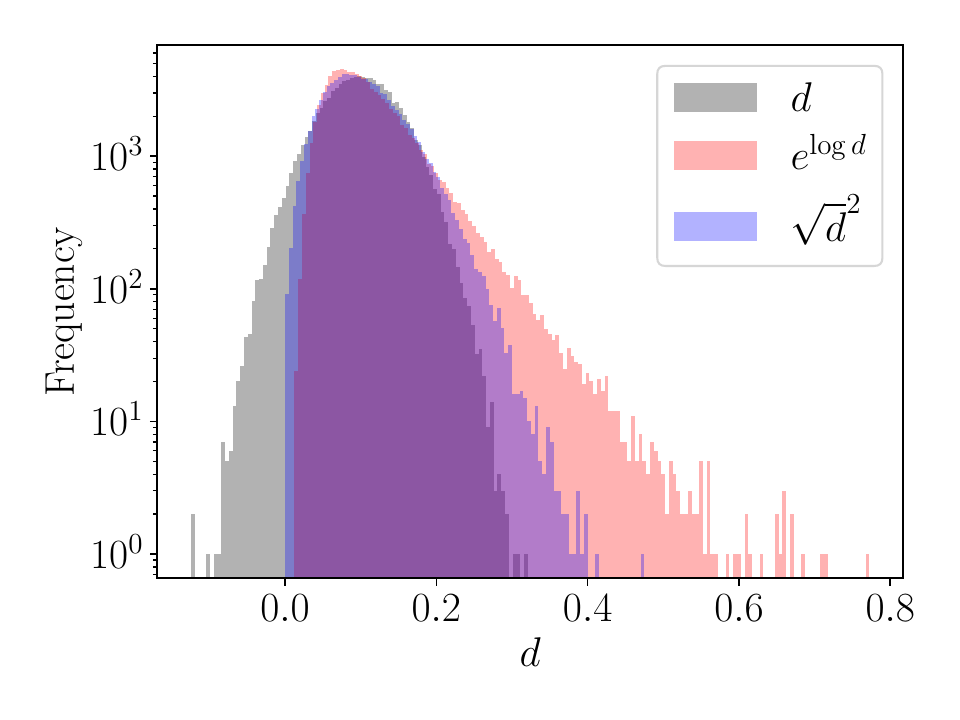}
  \caption{The Gaussian distribution (100000 samples) of $d_n=0.10(5)$ compared with logarithmic and square root distributions. Bottom: the same figure with a log scale on the $y$ axis for clarity.}
  \label{fig:dn_dist}
\end{figure}

Take as an example a prior for $d$ of 0.10(5), based on a typical size of effective amplitude $d^{\mathrm{eff}}\approx 0.1$ with enough uncertainty to allow for excited state amplitudes to be between 0 and roughly twice $d^{\mathrm{eff}}$, equivalent to choosing $A=1$ and $B=0.5$ in the above discussion, a reasonable assumption for an excited state amplitude prior.
Figure~\ref{fig:dn_dist} shows the distribution of a Gaussian $d=0.10(5)$ for reference, alongside the resulting distributions from taking a Gaussian for $\log(d)$ and $\sqrt{d}$. The bottom pane shows the same plot with a log $y$ axis for clarity. Of course, our Gaussian distribution is not suitable because it allows $d$ to be negative, but it does a good job of covering the positive values we might expect $d$ to be distributed within, so is a good comparison for the spread of values in the other distributions. We also see that the log distribution in particular has a large tail, allowing for values well beyond what we might want for a parameter we expect to be order 0.1. The square root distribution suffers less from this long tail. 
\begin{figure}
  \includegraphics[width=0.44\textwidth]{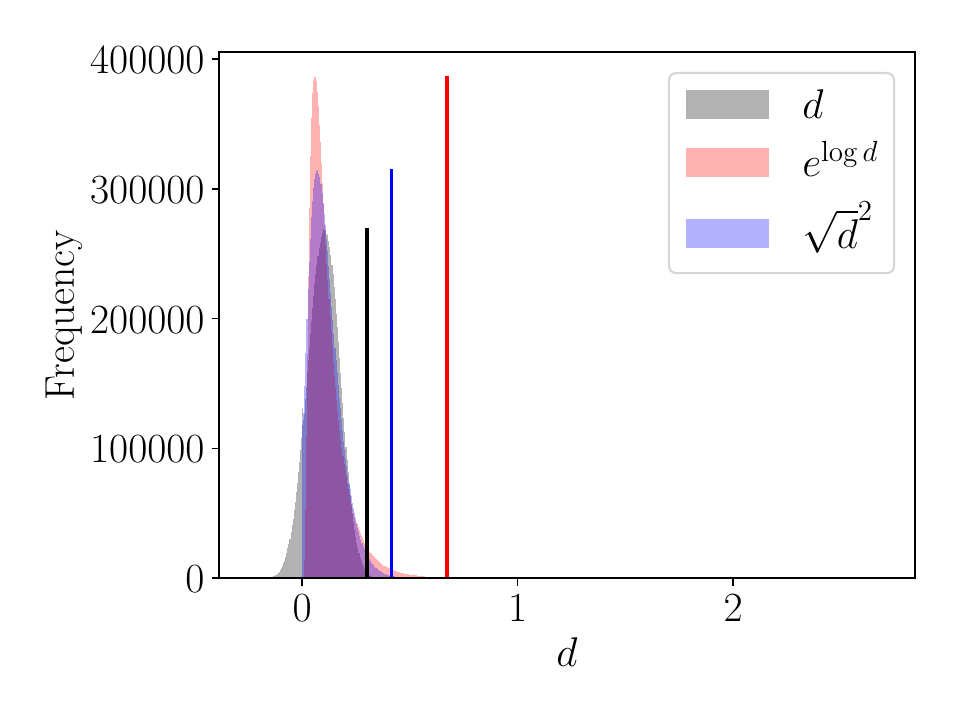}
  \includegraphics[width=0.44\textwidth]{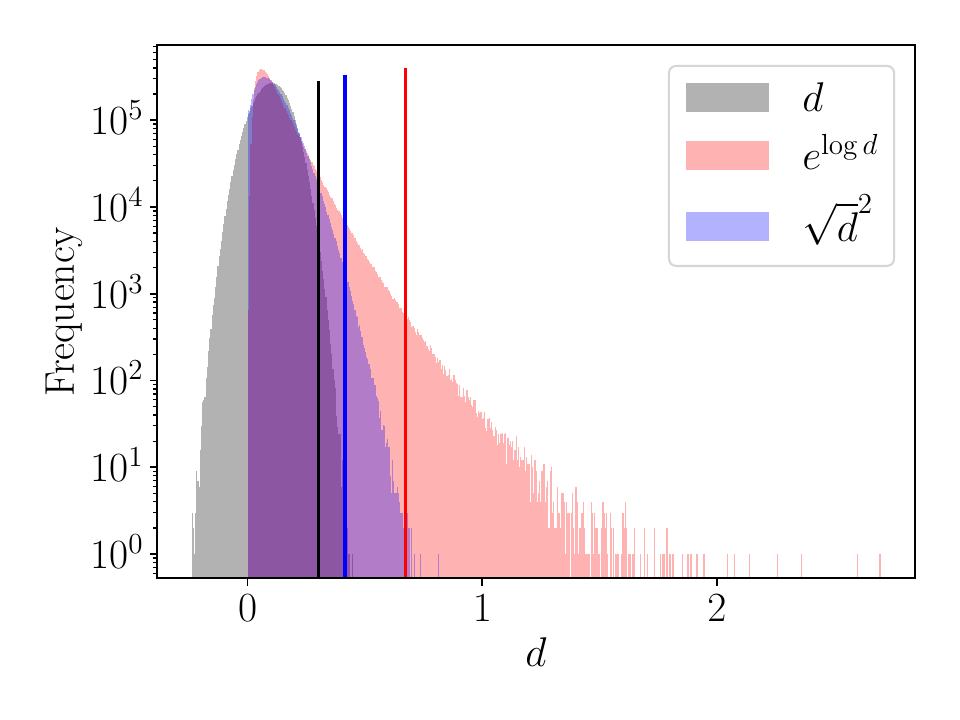}
  \caption{The Gaussian distribution of $d=0.10(5)$ compared with logarithmic and square root distributions when noise is applied (see text). The solid lines indicate the upper 99.86\% (+3$\sigma$) confidence point of the three distributions. Bottom: the same figure with a log scale on the $y$ axis for clairty.}
  \label{fig:dn_dist}
\end{figure}
This large tail is not a problem in and of itself - the fitter can still find the true value of the amplitude around 0.1, but it is greatly exacerbated by the addition of noise. We simulate this by drawing 100 central values at random for each of the Gaussian distributions (i.e. the Gaussian in $d$ space, Gaussian in log space and Gaussian in square root space). For each choice of central value, we generate the (100000 sample) distribution based on that new central value and the uncertainty. This is distribution of the `noisy' prior for that particular random choice of noise. We then convert each of these 100 noisy distributions back to $d$ space and add them all up. This gives an idea of the region of values explored by turning on prior noise. The figures include lines which give the 99.86\% upper confidence bound for the distributions (i.e. $3\sigma$ for the Gaussian). We see that the log distribution is much more biased towards large values than the Gaussian or square root. This means that, though it may do a good job of covering the expected range of $d$ in the fit, when the noise is turned on, it has a significant probability of choosing random values well beyond reasonable choices for $d$, which is undesireable. 

We find experimentally that square root distributions for amplitude priors behave much better under prior noise than log distributions. One issue arising from square root priors is that they can take negative values, which give the same value for $d$. Say for example that our prior $d=0.10(8)$, giving a $\sqrt{d}=0.32(13)$, if the true value of $d$ is $0.08$, say, then a $\sqrt{d}=0.28$ and $\sqrt{d}=-0.28$ value would both be acceptable solutions, from the $\chi^2_{\mathrm{data}}$ perspective. However, one if these is within $1\sigma$ of the prior, whilst the other is more than $4\sigma$ away, leading to large values for $\chi^2$ from the prior contribution, even though the contribution from the data is the same. This situation, where the fit finds the local negative $\chi^2$ minimum and not the global positive one is uncommon and easily rectified. Once the fit has converged, we take any negative values for $\sqrt{d}$ and multiply them by $-1$. We then rerun the fit with the previous result (including these new positive values) as the starting point (but providing the fitter with $p0=fit.pmean$) for a few iterations, causing the fit to rapidly converge on the positive result. 

A further way to improve the choice of distribution is to impose limits directly in the log or square root space. Let's say we are 95\% confident that our amplitude lies in the range $0.01\leftrightarrow 0.2$. Taking $d=0.10(5)$ say, which has a 95\% confidence interval of $0\leftrightarrow 0.2$ would be nice, but this prior allows negative values, so we choose to take the log or square root. The resulting $\log(d)=-2.30(50)$ or $\sqrt{d}=0.316(79)$, give 95\% confidence intervals in $d$ space of $0.04\leftrightarrow 0.27$ and $0.02\leftrightarrow 0.22$ respectively. These aren't bad approximations of the range we want, but they don't go so low as we'd like. Although 0.04 seems close to our desired 0.01, for example, in fact $d=0.01$ would correspond to $\log(d)=-4.6$, which is $4.6\sigma$ from the mean. The high end is also too large. This is exactly the problem demonstrated above.

One solution is to work out choices for $\log(d)$ or $\sqrt{d}$ which give a 66\% or 95\% or 99\% coverage for $d$ of our choosing. Take our above example with 95\% confidence that the excited state amplitudes lie in the range 0.01 to 0.2. We can then choose a prior in log space $x\pm \delta$, such that $e^{x-2\delta}=0.01$ and $e^{x+2\delta}=0.2$, or similarly in square root space.

\subsection{Example}\label{sec:example}
Let's put everything we've discussed so far in to an example. We'll start with some raw two-point correlators, and plot the effective masses and amplitudes. We use the following equations, which use $C_2(t\pm 2)$ rather than $C_2(t\pm1)$ because of the oscillating states in our correlators,
\begin{equation}\label{Eq:Meff}
  aM_{\text{eff}}(t) = \frac{1}{2}\cosh^{-1}\Bigg(\frac{C_2(t-2)+C_2(t+2)}{2C_2(t)}\Bigg),
\end{equation}
\begin{equation}\label{Eq:Aeff}
d_{\text{eff}}^2(t)  = \frac{C_2(t)}{e^{-M_{\text{eff}}t}+e^{-M_{\text{eff}}(N_t-t)}}.
\end{equation}
We can then plot $aM_{\mathrm{eff}}(t)$ and $d_{\mathrm{eff}}^2(t)$ as a function of $t$ and eyeball a plateau. Better still, for large numbers of correlators, we can write code to look for plateaus, performing a rolling average over two or four adjacent points (to average oscillations), and looking for the region where this average changes the least between time slices. The code could also take into account the uncertainty in the average, in order to avoid settling on a region which is pure noise and happens to average to a consistent value. It's a matter of choice whether to work with $d_{\mathrm{eff}}^2(t)$, as we prefer to, or to take the square root before finding the plateau.   
\begin{figure}
  \includegraphics[width=0.44\textwidth]{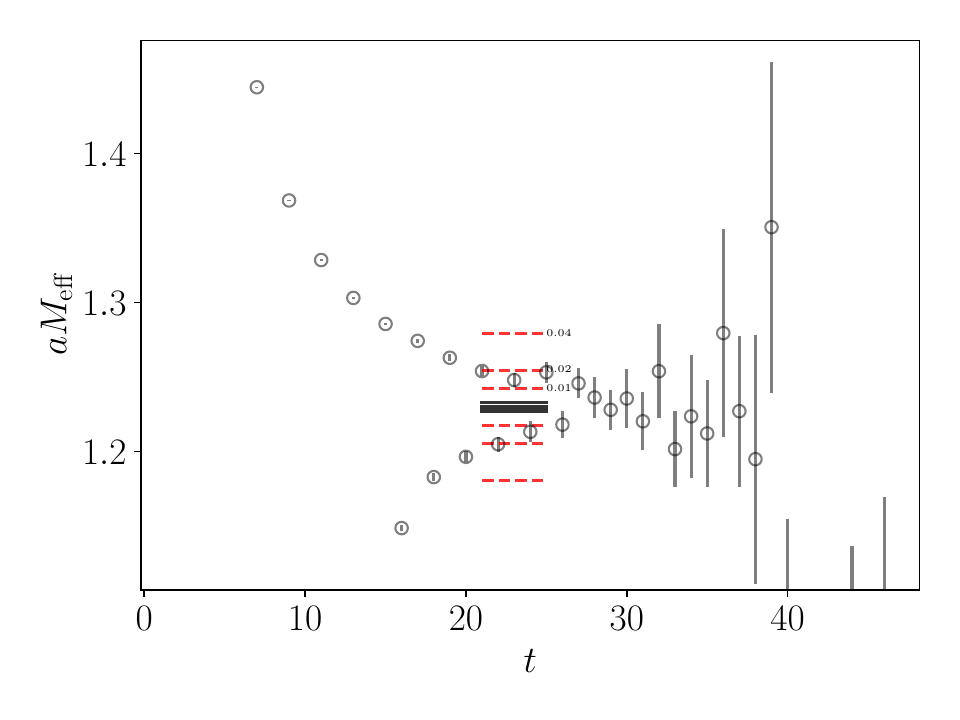}
  \includegraphics[width=0.44\textwidth]{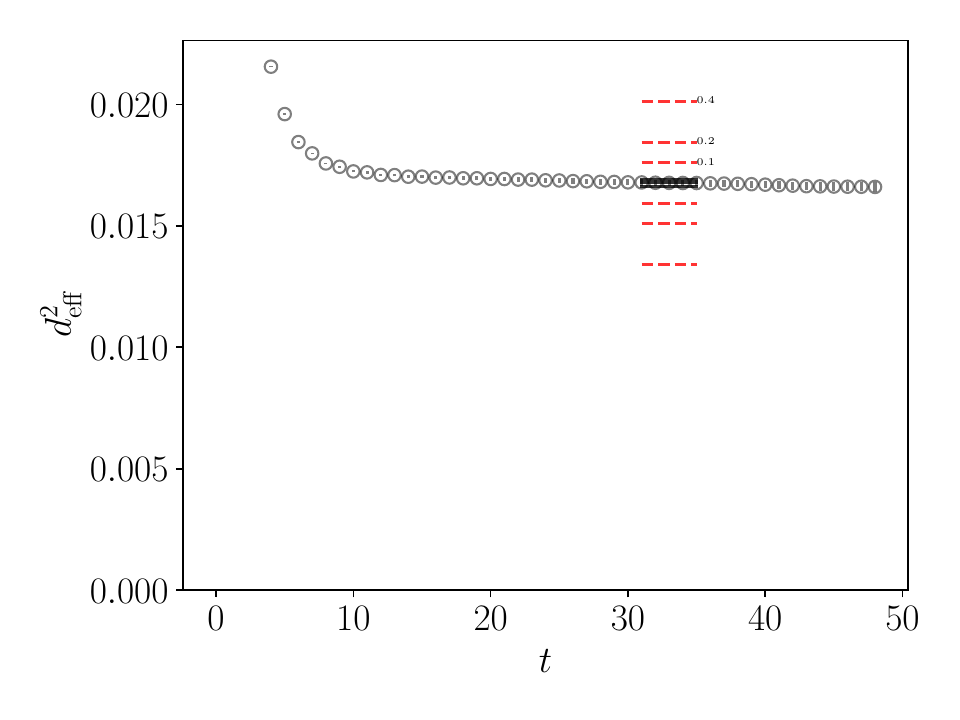}
  \caption{$aM_{\mathrm{eff}}(t)$ (top) and $d_{\mathrm{eff}}^2(t)$ (bottom) for a choice of two correlators. The black band indicates the range of $t$ values which the code identified as being the plateau and the depth is the error bar on their average. Red dotted lines indicate relative uncertainties on this value 1\%, 2\% and 4\% for $aM_{\mathrm{eff}}(t)$, and 10\%, 20\% and 40\% for $d_{\mathrm{eff}}(t)$. Note that the latter uncertainties applied to the unsquared value, as this is what our prior will be.}
  \label{fig:M_a_eff}
\end{figure}

Figure~\ref{fig:M_a_eff} shows a example of an effective mass and effective amplitude plot for different correlators. We see the values approach a plateau as $t$ increases. Oscillations are present in both figures, but we deliberately chose an effective mass plot with particularly large oscillations and noise in order to demonstrate the procedure. The $t$ extent of the black band indicates the $t$ values which the code finds as being the plateau, and the depth of the band indicates the uncertainty on this value. The red dotted lines are to guide the eye. They mark 1\%, 2\% and 4\% for $aM_{\mathrm{eff}}(t)$, and 10\%, 20\% and 40\% for $d_{\mathrm{eff}}(t)$ error bars. We note that in the case of the amplitude, these error bars indicate the specified uncertainty on the amplitude itself, not on the amplitude squared, which is what is plotted.

Taking the information from these plots, we can determine priors for the non-oscillating ground state energy and amplitude, as well as reasonable uncertainties on these priors. For example, in the case of the mass we have a central value of 1.227 and we can take an uncertainty of 4\%, which our plot indicates is conservative, giving a prior $P[aE_0^{\mathrm{n}}]=1.227(50)$. We can repeat this for the amplitude, taking an uncertainty of 20\%, say. We choose a percentage uncertainty which is reasonable for all the correlators in our fit, and this leaves us with just two values (the percentage uncertainty on all ground state non-oscillating masses, and all ground state non-oscillating amplitudes) to vary in our stability analysis.

For all excited state (oscillating and non-oscillating) energy \textit{splittings} ($aE_{i\neq0}$ in \textit{corrfitter}), we usually take $\Lambda_{\mathrm{QCD}}\pm\Lambda_{\mathrm{QCD}}/2$. We can of course play with this, use log or square root priors, and set confidence limits, as discussed above. In general it's worth thinking about the relationship between $d_i$ and $E_i$. For example, if we give $d_i$ very broad priors, which allow very small values, and we give narrow priors for $E_i$, then the fitter may happily fit energies which don't appear in the spectrum, with very small amplitudes. This is best avoided, but in principle it shouldn't mess up the ground state energy, which is usually what we're actually interested in.

For the ground state oscillating energy, we may take as the prior $P[aE_0^{\mathrm{o}}]=P[aE_0^{\mathrm{n}}] + a\Lambda_{\mathrm{QCD}}\pm a\Lambda_{\mathrm{QCD}}/2$. This means we slightly adjust $P[aE_0^{\mathrm{o}}]$ when we experiment with the percentage uncertainty on $P[aE_0^{\mathrm{n}}]$, but we still have a large extra uncertainty reflecting our prior knowledge that this splitting is of order $a\Lambda_{\mathrm{QCD}}$.

For our excited state amplitudes, and oscillating ground state amplitude, we can use the log or square root priors discussed above, with the prior knowledge that the amplitudes are likely to be of the order of the ground state $P[d_0^{\mathrm{n}}]=d^{\mathrm{eff}}(1\pm\mathrm{\% error})$ or smaller, assuming our operators are optimised for ground state overlap. It's not usually worth taking the lower bound on these priors too low. If the ground state amplitude is 0.1, say, and we might expect to resolve it with a 1\% uncertainty\footnote{The rough size of this can be established from the effective amplitude uncertainty - the actual uncertainty found on $d^{\mathrm{eff}}$, not the conservative \% we use for the prior.}, then there's no point allowing priors for excited states to be much smaller than 0.001, as, especially when multiplied by heavier decaying exponentials, these will be lost in the noise. 

Finally, we might choose to boost or reduce the size of oscillating amplitudes relative to non-oscillating ones, based on log plots of the correlators. These observations will vary from situation to situation, but for $H\to K$ with a HISQ action, we observe that oscillations are small in kaons, particularly at low or zero momentum, and they are larger in $H$ mesons, particularly if the have `non-Goldstone' spin-taste.  
\begin{figure}
  \includegraphics[width=0.44\textwidth]{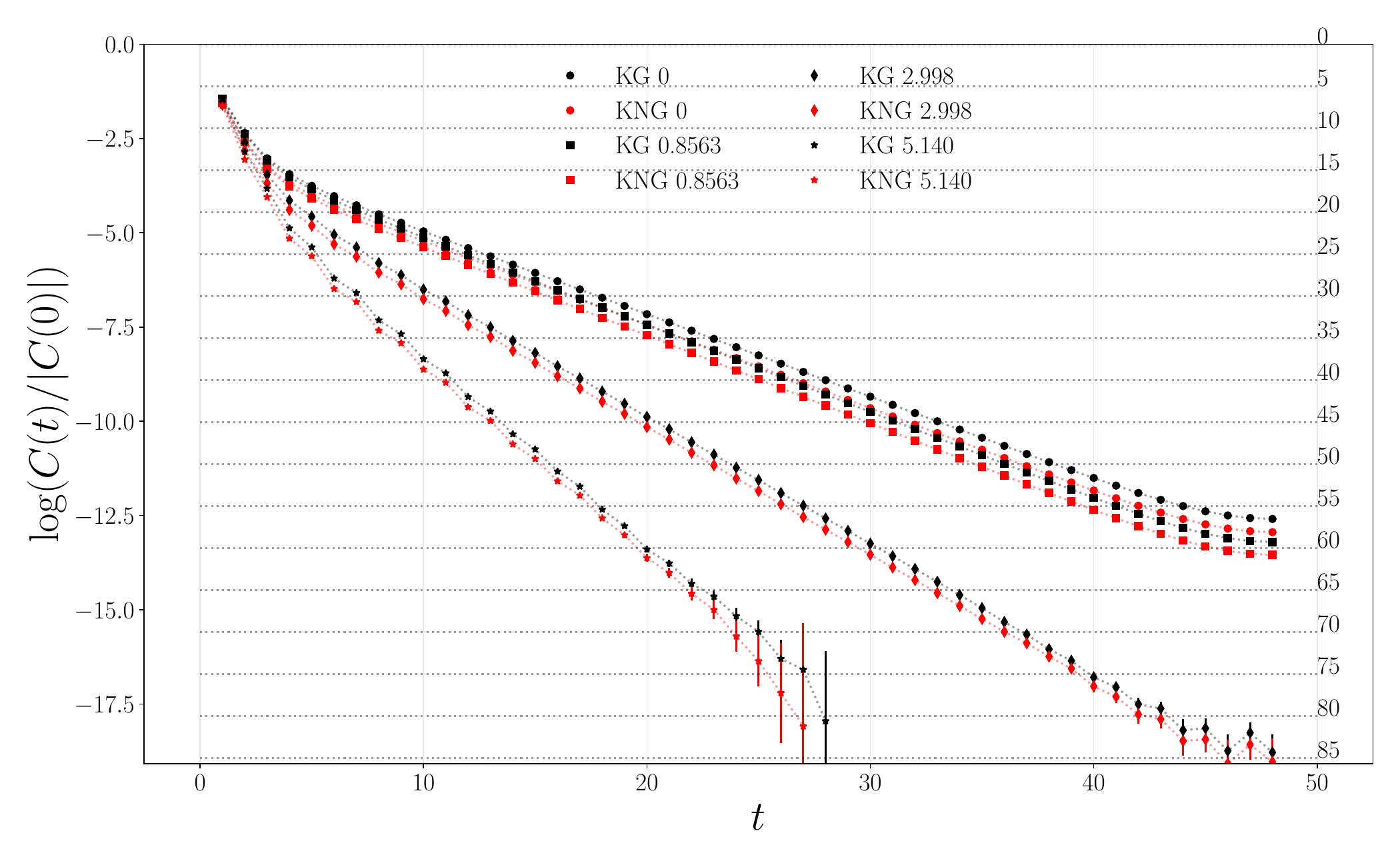}
  \includegraphics[width=0.44\textwidth]{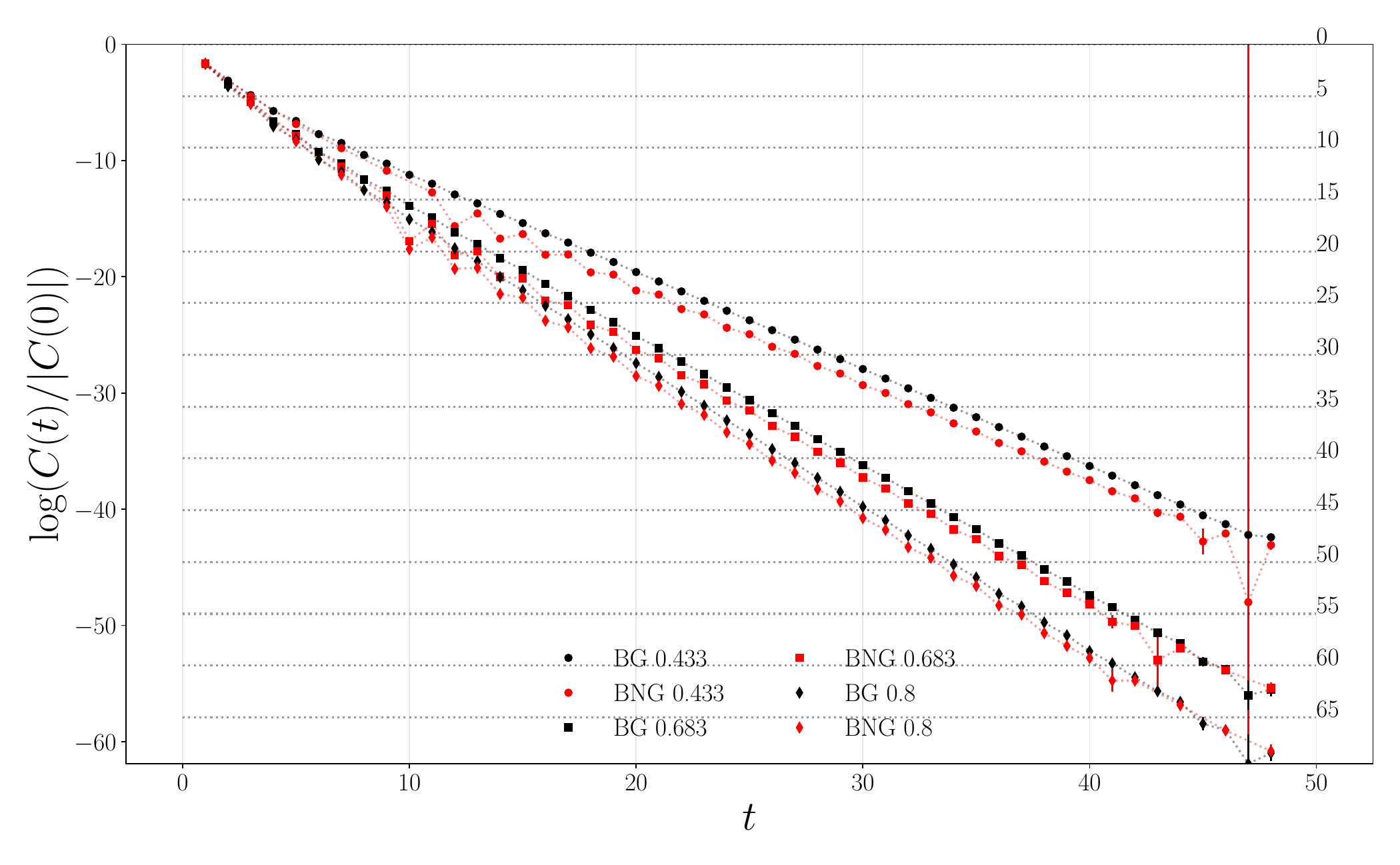}
  \caption{The log of $K$ (top) and $H$ (bottom) correlators, for increasing $m_h$ values and momenta (twists). The black points show Goldstone (`G') spin-taste, and the red non-Goldstone (`NG').}
  \label{fig:log_C}
\end{figure}
Figure~\ref{fig:log_C} demonstrates this. The plots show the log of $K$ (top) and $H$ (bottom) correlators, for increasing $m_h$ values and momenta (twists). The black points show Goldstone (`G') spin-taste, and the red non-Goldstone (`NG'). We can clearly see oscillations in the $K$ with the largest twist, as well as in the $H$ in the non-Goldstone case. We may choose to broaden the range of the priors $P[d_{i}]$ in cases where we can see that the oscillations are large.   

We now have priors for all the two-point parameters, and we should move on to the three-points. Firstly, we calculate the effective $J^{\mathrm{nn,eff}}_{00}$, using one of two equations,
\begin{equation}\label{eq:Jeff}
  J^{\mathrm{nn,eff}}_{00}(t, T) = \frac{C_3(t, T)}{C^{M_1}_2(t)C^{M_2}_2(T-t)} d_{\rm eff}^{M_1}d_{\rm eff}^{M_2},
\end{equation}
or
\begin{equation}\label{eq:Jeff_Meff}
  J^{\mathrm{nn,eff}}_{00}(t, T) = \frac{C_3(t, T)}{d_{\rm eff}^{M_1}d_{\rm eff}^{M_2}e^{-M_{\mathrm{eff}}^{M_1}t} e^{-M_{\mathrm{eff}}^{M_2}(T-t)}},
\end{equation}
where $M_1$ and $M_2$ are $H$ and $K$ as before, and $M_{\mathrm{eff}}$ are their respective masses. Both of these methods should give the same plateau, but we find in our case that they approach this plateau in very different ways. 
\begin{figure}
  \includegraphics[width=0.44\textwidth]{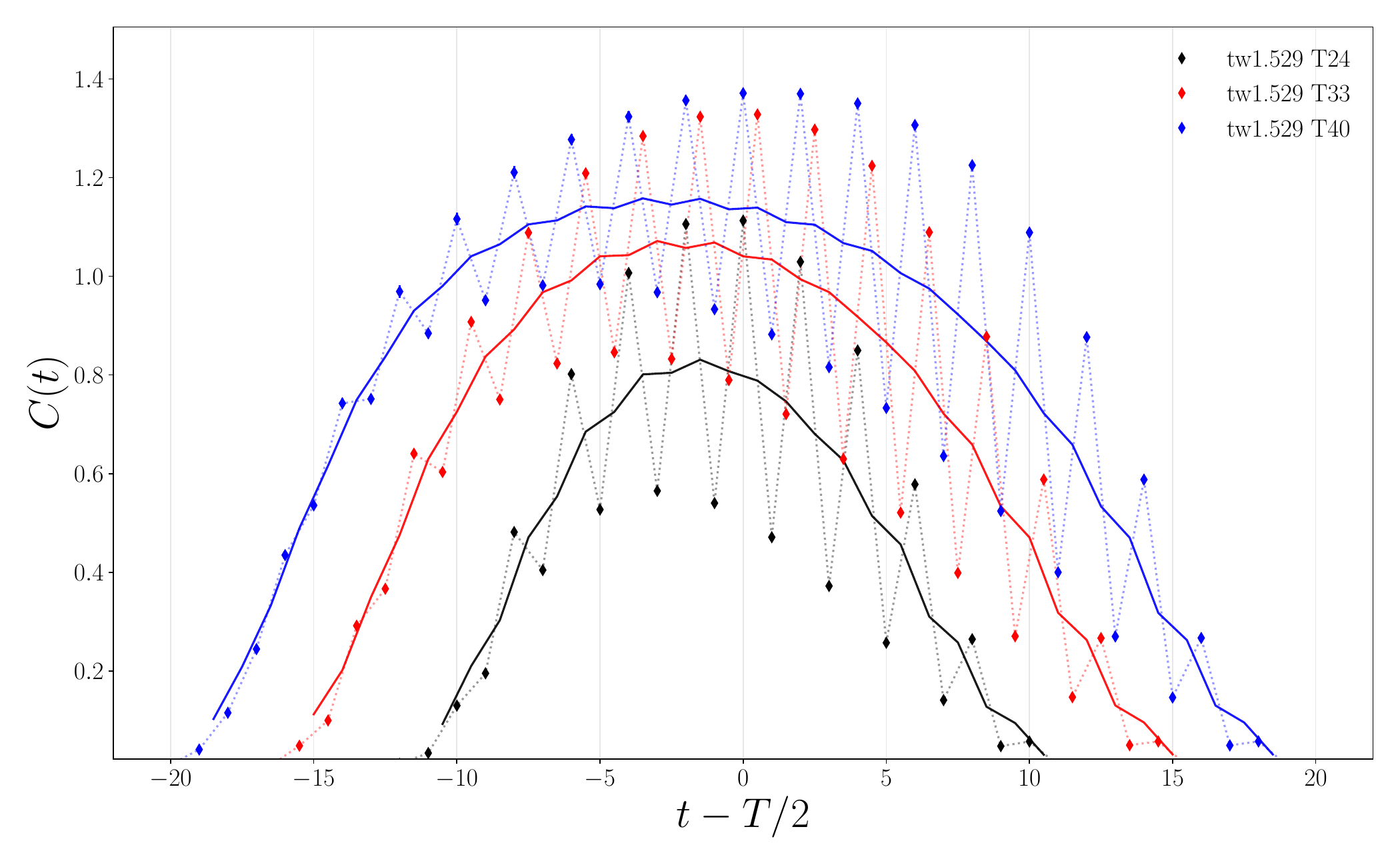}
  \includegraphics[width=0.44\textwidth]{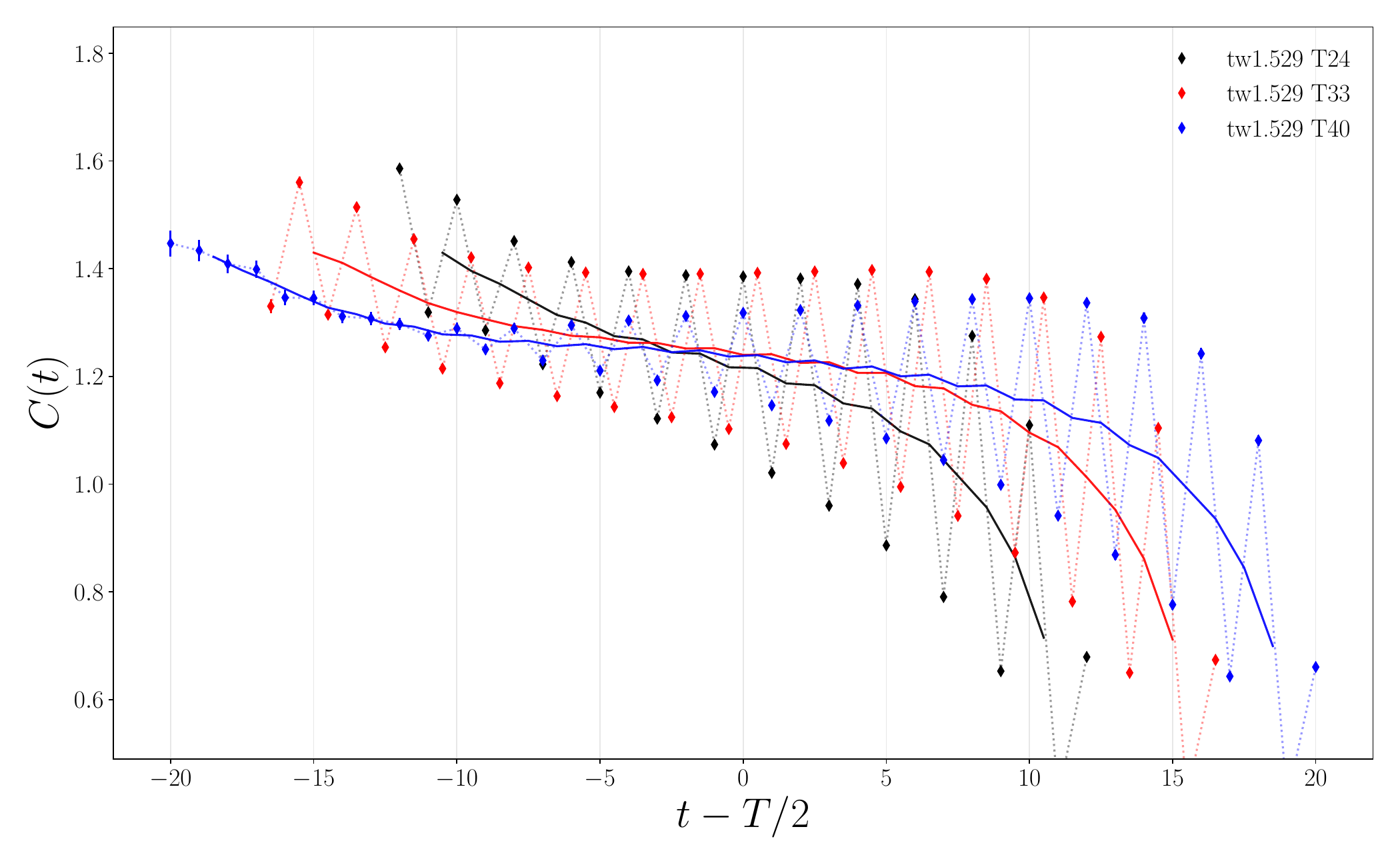}
  \caption{$S^{\mathrm{nn,eff}}_{00}(t, T)$ for three different $T$ values, using Eq.~\eqref{eq:Jeff} (top) and Eq.~\eqref{eq:Jeff_Meff} (bottom). The lines indicate a rolling average over pairs of data points.}
  \label{fig:Jeff_S} 
\end{figure}
This is clear from Figure~\ref{fig:Jeff_S}, which shows the two approaches applied to a scalar current insertion, for three different $T$ values. Both figures seem to suggest a value in the region of 1.2, but the method using Eq.~\eqref{eq:Jeff} approaches this from below, getting larger with increasing $T$, whilst the method from Eq.~\eqref{eq:Jeff_Meff} seems to give a more consistent value across different $T$ choices. Oscillations are very obvious in both plots. Whilst either could in principle be used to estimate $J^{\mathrm{nn,eff}}_{00}(t, T)$, different techniques may be considered. For example, the top plot might call for finding the plateau in $J^{\mathrm{nn,eff}}_{00}(t, T)$ for each $T$ value, and choosing the largest of these (or equivalently the largest $T$), whilst the lower plot could be used to take an average of all $T$ values. In either case, it is clear that a large uncertainty, of the order of at least 30\%, is required, given the variation in the plots.
\begin{figure}
  \includegraphics[width=0.44\textwidth]{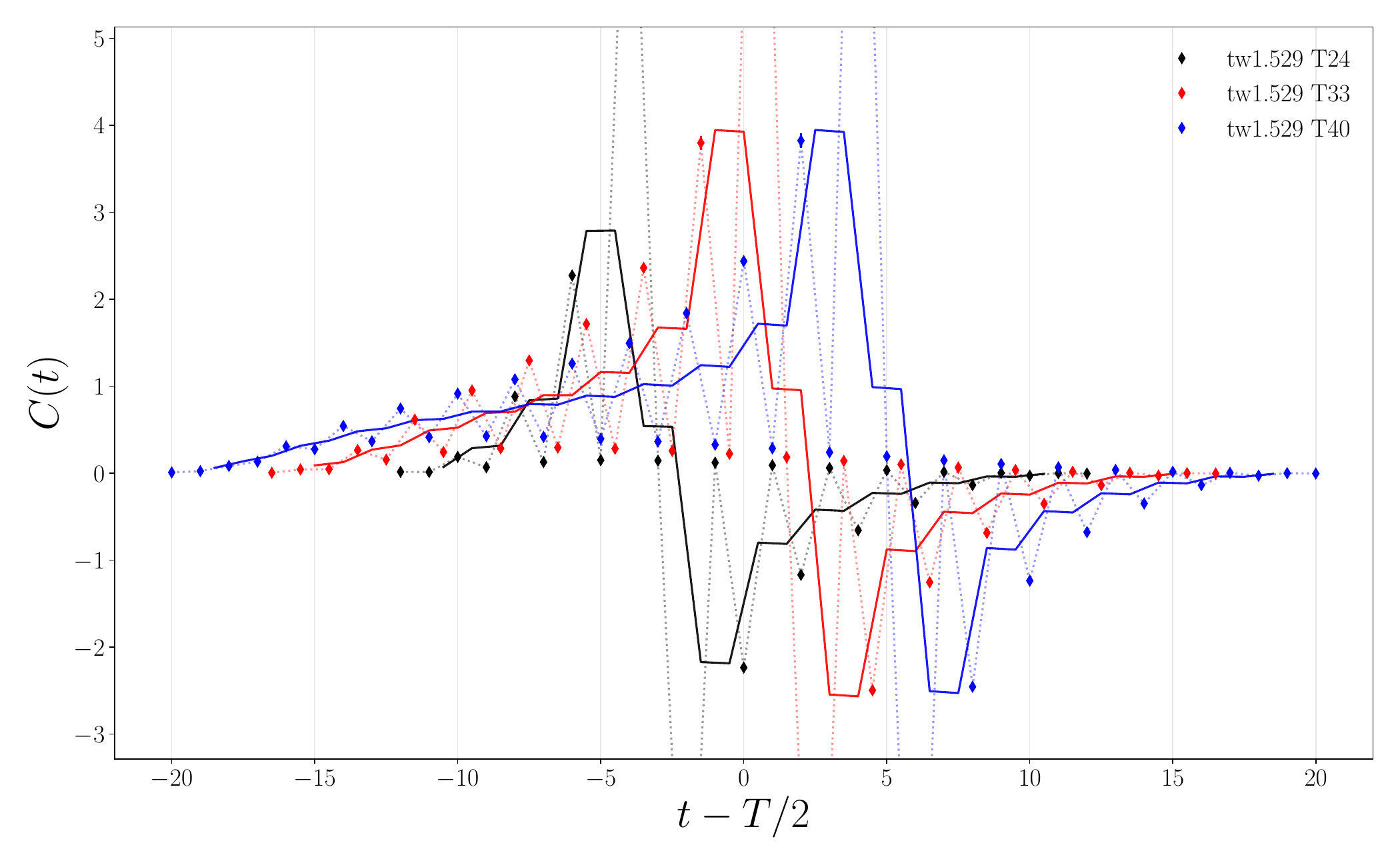}
  \includegraphics[width=0.44\textwidth]{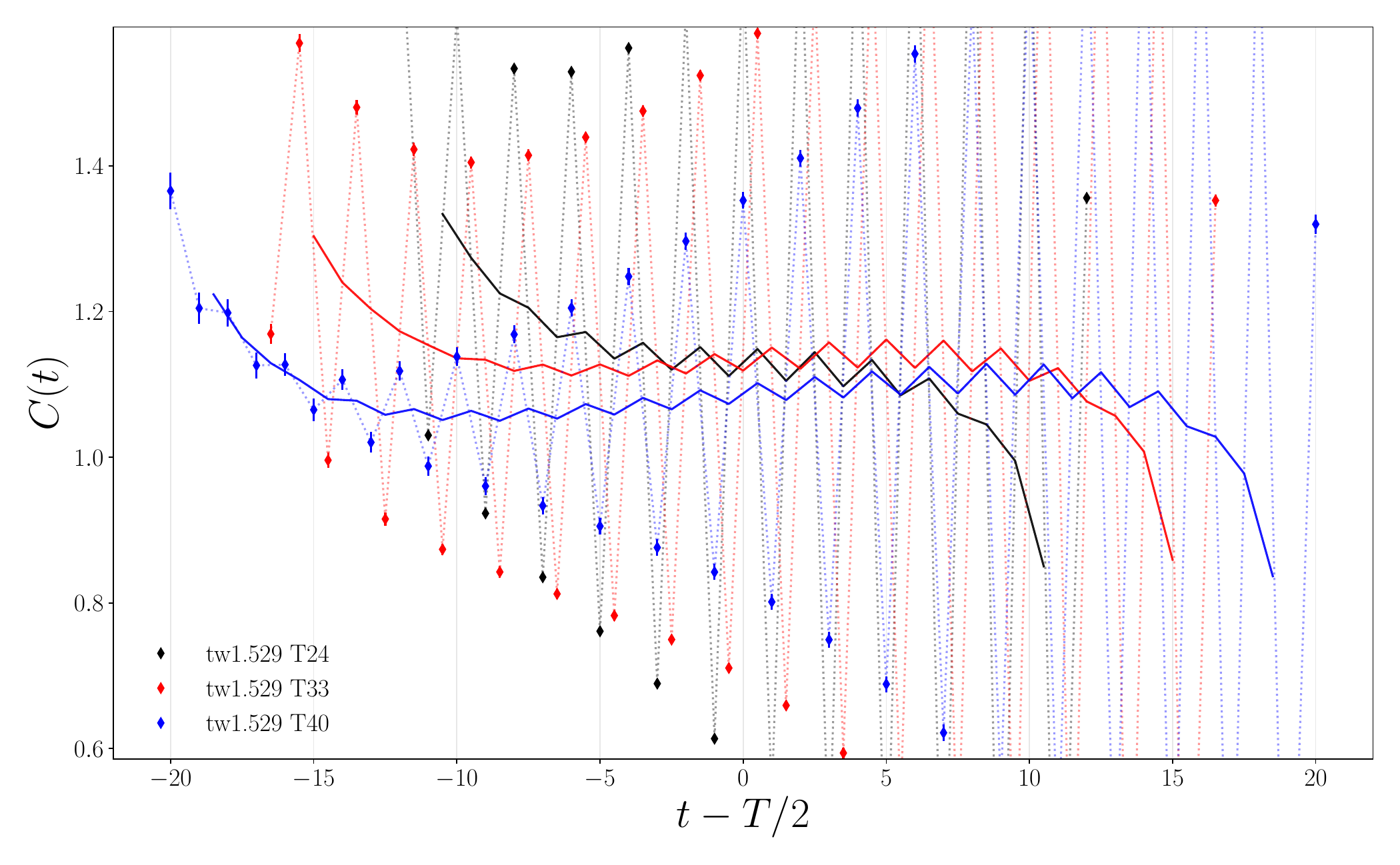}
  \caption{$V^{\mathrm{nn,eff}}_{00}(t, T)$ for three different $T$ values, using Eq.~\eqref{eq:Jeff} (top) and Eq.~\eqref{eq:Jeff_Meff} (bottom). The lines indicate a rolling average over pairs of data points.}
  \label{fig:Jeff_V}
\end{figure}

Figure~\ref{fig:Jeff_V} is a similar plot to Figure~\ref{fig:Jeff_S}, drawn from the same data set, only with a vector current insertion. We can see that the behaviour for the Eq.~\eqref{eq:Jeff} method is wildly different to the scalar case. This is common across all of our data for different masses and momenta. Whilst it is still possible to find a reasonable estimate from the Eq.~\eqref{eq:Jeff} in this case, by performing a rolling average over four data points and looking for a plateau, it is clear that Eq.~\eqref{eq:Jeff_Meff} is likely to give a much more reliable value for $J^{\mathrm{nn,eff}}_{00}(t, T)$ in this case. 

Taking these plots, we might use the prior $P[J^{\mathrm{nn}}_{00}] = J^{\mathrm{nn,eff}}_{00}\pm 0.5 J^{\mathrm{nn,eff}}_{00}$, for example, where we vary the percentage error in our stability analysis. As discussed above, we then base our priors for non-ground state and ground state oscillating terms off the reasonable assumption that they will be of the same order of magnitude, and probably smaller. $P[J^{kl\neq\mathrm{nn}}_{00}] = 0 \pm 1.0 J^{\mathrm{nn,eff}}_{00}$ and $P[J^{kl}_{ij\neq 00}] = 0 \pm 0.5 J^{\mathrm{nn,eff}}_{00}$, for example, where again, we can vary the percentage uncertainty across all these priors simultaneously by adjusting the (in this case) 1.0 and 0.5 choices in our stability analysis.

This concludes our prior selection. We have hopefully achieved our goal of choosing sensible priors that accurately reflect our prior knowledge for all parameters, whilst simultaneously reducing the chosen parameters to adjust in a stability analysis to a manageable number.

\section{Improving statistics}
Once we've chosen our priors, the next objective is to get as much information from our data as possible. This means maximising the samples, $N_{s}$, and minimising the data points we use, $N_{G}$. Both of these act to reduce the required SVD cut, and so improve the precision, and the latter also reduces the size of the covariance matrix, thus reducing the complexity of the fit and speeding everything up. As a balance to this, reducing the data points also pushes up the $\chi^2$. 
\subsection{Binning over $t_0$}
For any given ensemble, we have $n_{\mathrm{cfg}}$ gluon field configurations to sample. These configurations should be statistically independent samples, assuming they have been generated and tested correctly. Within each ensemble, we typically calculate our correlators with $n_{\mathrm{src}}$ different source times $t_0$, evenly spaced, with a varying start point. The number of time sources per ensemble in our particular data set is 16.

As previously noted, \textit{corrfitter}~\cite{corrfitter} takes $N_s$ uncorrelated samples of our correlator and builds an average, a covariance matrix, and an appropriate SVD cut.

Whilst we know that individual configurations are uncorrelated, it is likely, or at least possible, that correlators starting at different source times $t_0$ on a given ensembles are correlated. For this reason, we must average over (bin) our $n_{\mathrm{src}}$ samples on each ensemble before generating our covariance matrix. This ensures the covariance matrix is built using uncorrelated samples, but means we are limited to $N_s=n_{\mathrm{cfg}}$. However, it is worth checking whether or not this is indeed the case, as if different $t_0$ correlators turn out to be sufficiently uncorrelated, we can treat them as independent, which would increase our sample number to $N_s=n_{\mathrm{cfg}}\times n_{\mathrm{src}}$, thereby reducing our SVD cut and final fit uncertainties. We could also partially bin, for example binning over 8 sources and not 16, giving $N_s=2n_{\mathrm{cfg}}$.

\begin{figure}
  \includegraphics[width=0.48\textwidth]{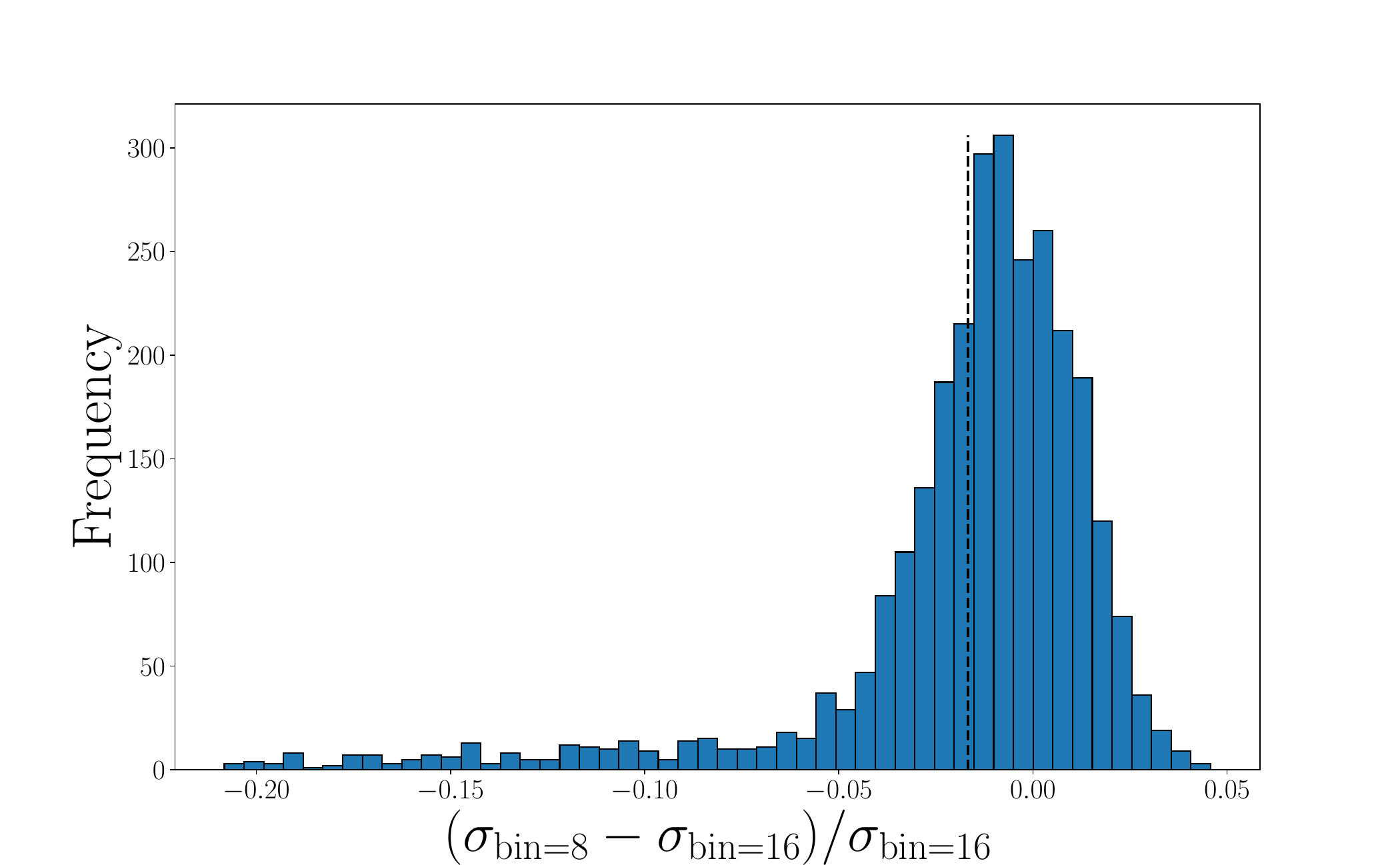}
  \includegraphics[width=0.48\textwidth]{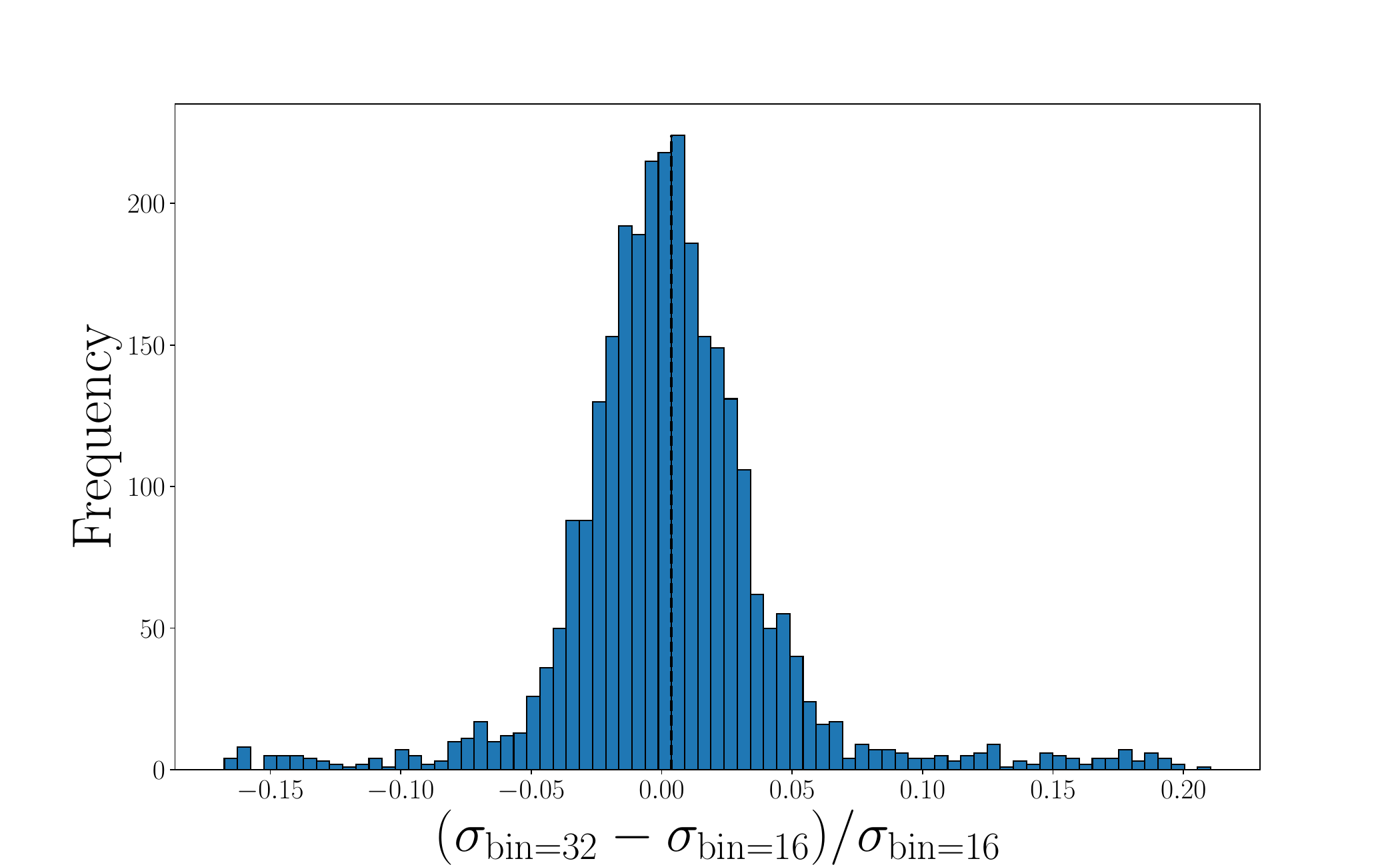}
  \caption{Relative changes in uncertainty on all correlator time slices from the default binning of 16 to a binning of 8 (top) or 32 (bottom).}
  \label{fig:Cp_binning}
\end{figure}

One method to establish how correlated the different samples are involves binning the data into different sized bins, with binning over all $n_{\mathrm{src}}$ sources being the default, which we know offers uncorrelated samples. For each bin size, we average the data and generate the covariance matrix. Then we take each timeslice of each correlator and compare the relative change in the uncertainties from the default case. If the samples in the new binning regime are uncorrelated, we would expect the relative change in uncertainty to form a Gaussian distribution, with a mean of zero. In the case that they are correlated, we would expect the average standard deviation to be smaller than in the uncorrelated case. We give an example of the results from this in Fig.~\ref{fig:Cp_binning} for one particular ensemble. 

The upper pane shows the relative change in size of uncertainties on correlator time slices between the default binning of 16 and a reduced binning of 8. We see that the average reduction in the uncertainty is about 2\%, but the distribution is slightly skewed, indicating that the sources are somewhat correlated. On the other hand, if we increase the binning to 32, we see a near Gaussian distribution, with a mean very close to 0, (and relatively symmetric tails). This seems to indicate that the default binning of 16 is a good choice to obtain uncorrelated samples.

We can take this one step further, however, as this distribution might hide different behaviour for different correlators.
\begin{figure}
  \includegraphics[width=0.48\textwidth]{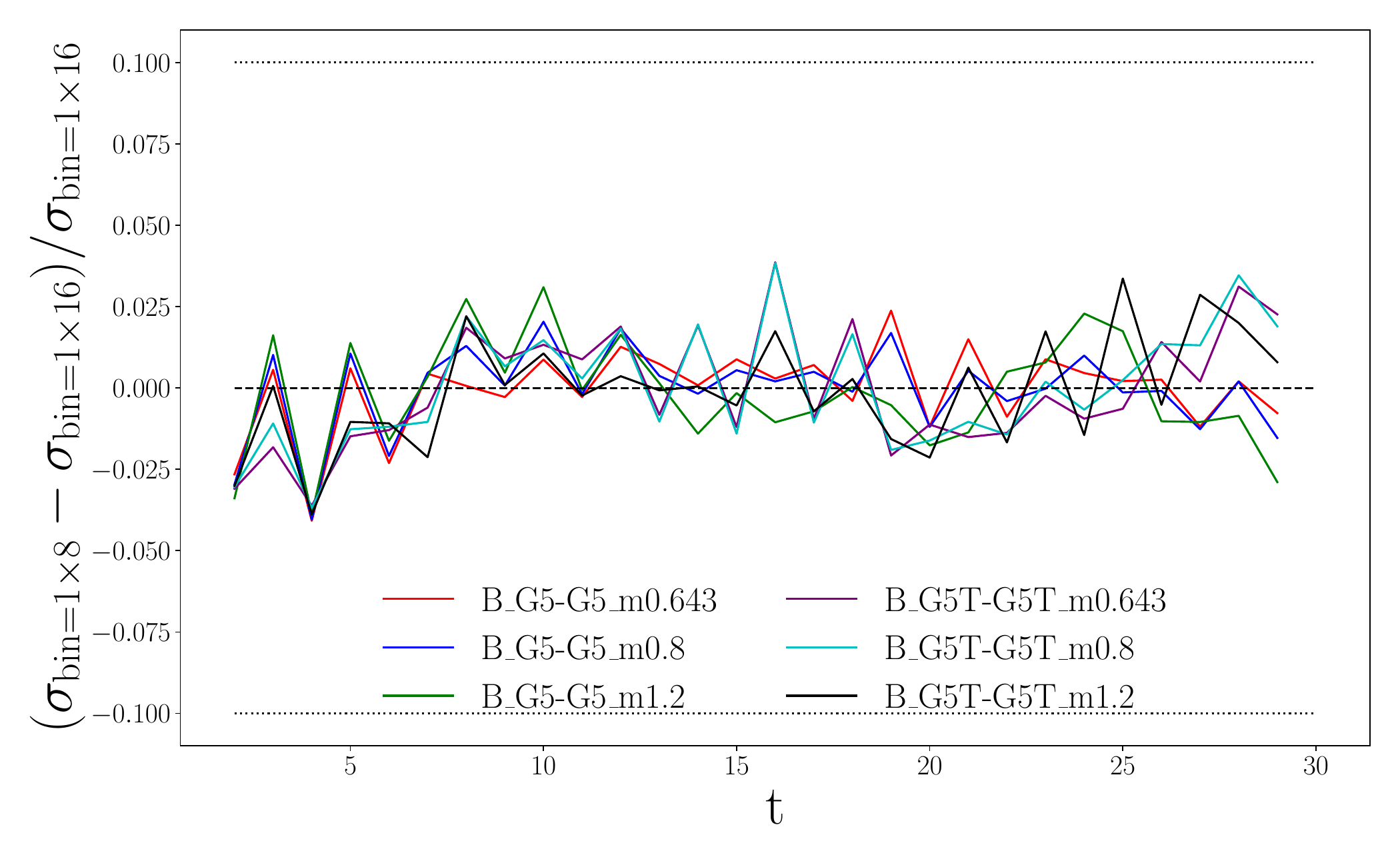}
  \includegraphics[width=0.48\textwidth]{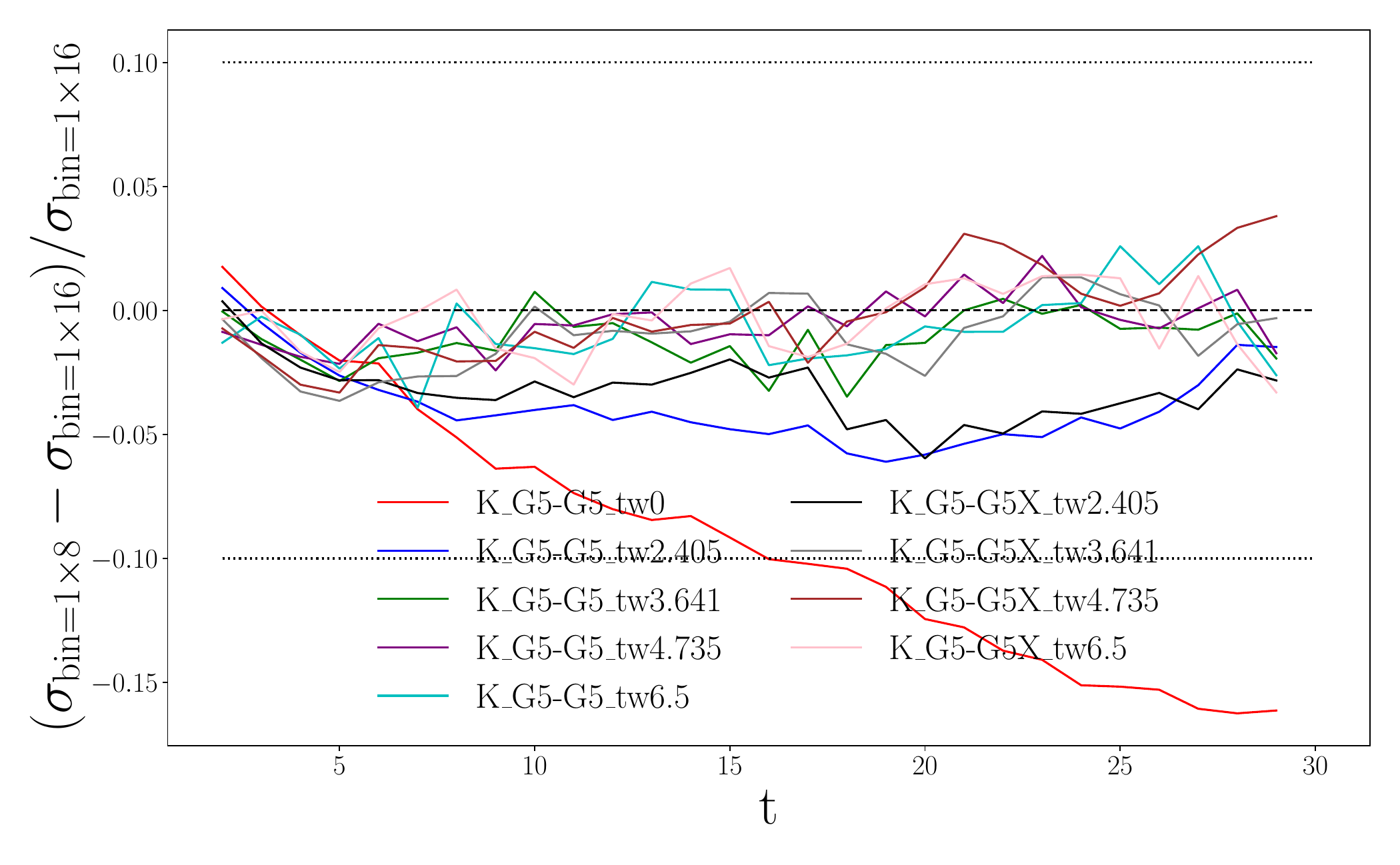}
  \includegraphics[width=0.48\textwidth]{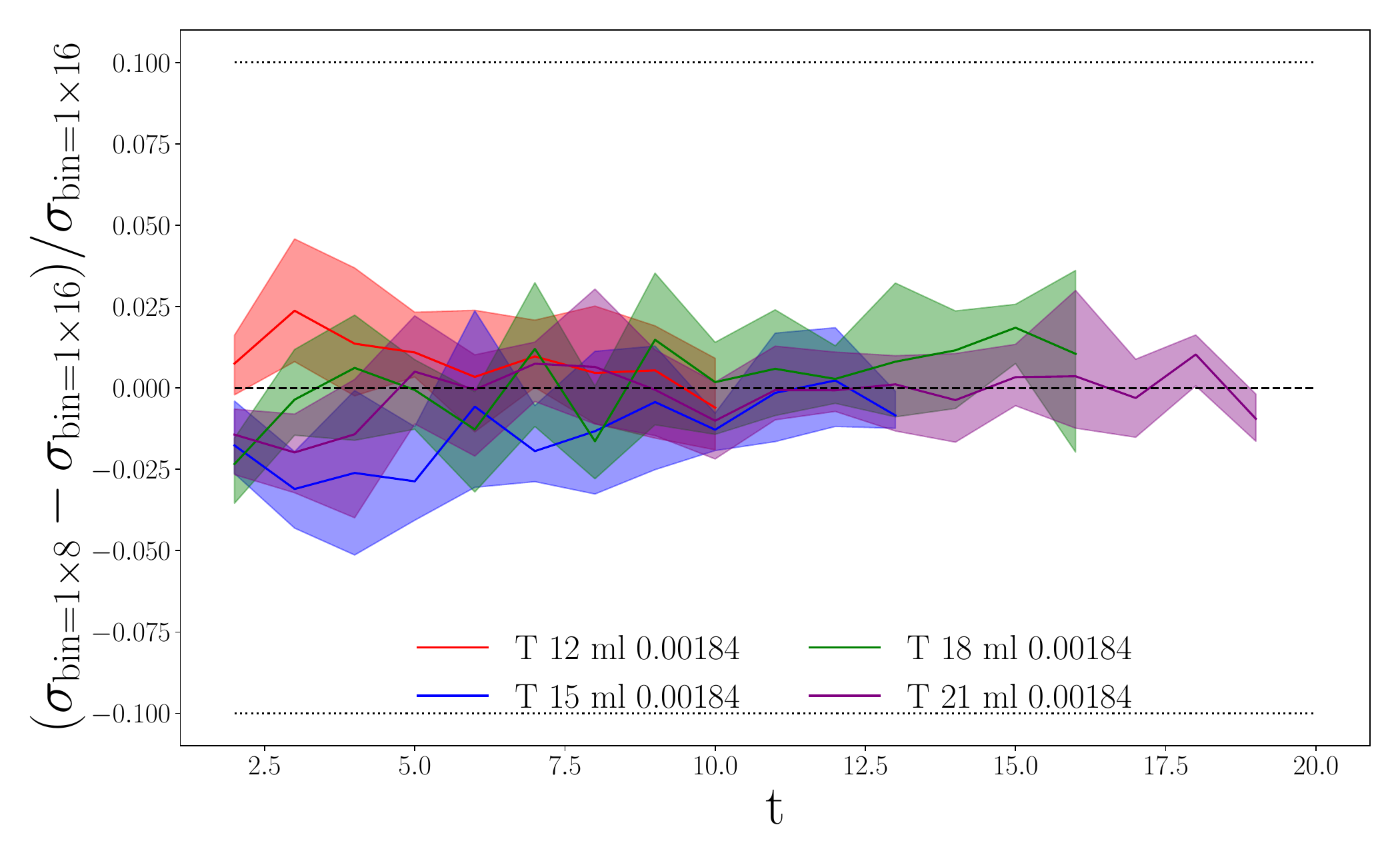}
  \caption{Relative changes in uncertainty on all correlator time slices from the default binning of 16 to a binning of 8. Broken down into two-points $H$ (top) and $K$ (middle) and the three point tensor for all four $T$ values (bottom). See text for more details. Dotted lines indicate a relative change of $\pm10\%$}
  \label{fig:Cp_binning_breakdown}
\end{figure}
Figure~\ref{fig:Cp_binning_breakdown} shows the same ensemble, again looking at the difference between a binning of 16 and 8. The top pane shows all of the $H$ meson correlators, labelled according to their spin taste (G5-G5 is the `Goldstone' and G5T-G5T the `non-Goldstone') and heavy mass\footnote{The precise meaning of these labels is not relevant to the discussion.}. For each time $t$, the relative change in uncertainty on the correlator is plotted. We see that for all the correlators in this case, the change is within $\approx\pm2.5\%$ and seems to be relatively evenly distributed about 0. This suggests that for the $H$ correlators on this ensemble, binning over 8 sources instead of 16 still gives us independent samples.

For the $K$ mesons in the middle pane of Figure~\ref{fig:Cp_binning_breakdown}, the story is slightly different. Again we plot the relative change in uncertainty for each timeslice of each correlator. This time we have spin tastes G5-G5 and G5-G5X and a variety of twists, labelled `tw'. In this case, there is very different behaviour for different correlators. Whilst those with larger twist behave similarly to the $H$ mesons above, indicating no correlations between samples, the small twist $K$s, and in particular the twist 0 kaon (shown in red), appear to be much more strongly dependent on the binning, showing a reduction of uncertainty of up to 15\%.

The bottom pane of Figure~\ref{fig:Cp_binning_breakdown} is a similar plot, showing the tensor current insertion for the four different $T$ choices. In this case, we have many such correlators for each $T$, one for each mass and twist combination. We display them compactly by plotting the average as the coloured line, with a band extending to the minimum and maximum. We see that all changes here lie within $\pm 5\%$. We can make similar plots for the other current insertions.

So should we choose a binning of 8 or 16? Here there is a decision to be made about what is an acceptable (i.e. unimportant) relative change in uncertainty. For the sake of this example let's say that we are happy with any change less than 10\%, which seems a reasonable position. We have plotted the $\pm10\%$ lines on each figure to guide the eye. We can see that for our case, we would be very happy to bin over 8 sources instead of 16, but for the twist 0 Goldstone $K$ (in red), which shows deviation beyond our chosen 10\% limit for $t>15$. One solution to this would be just to stick with the default binning over 16 $t_0$s, but another would be to say that we can bin over 8, but in that case we shall discard data from this correlator for $t>15$. As with everything discussed so far, the recommendation is to try both of these as part of a comprehensive stability analysis.   
\subsection{Choosing $t_{\mathrm{min}}$ and $N_{\mathrm{exp}}$}\label{sec:tmin}
Another choice to be made in our fits is the $t$ range $t_{\mathrm{min}}\leq t\leq t_{\mathrm{max}}$ over which we fit the two, and three-point correlators (Equations~\eqref{eq:2ptcorr} and ~\eqref{eq:3ptcorr}). Because the correlators are an infinite sum, the $t_{\mathrm{min}}$ value we choose is necessarily linked to the number of exponentials $N_{\mathrm{exp}}$ we must fit. In the limit $t_{\mathrm{min}}=0$, $N_{\mathrm{exp}}=\infty$, but as we increase $t_{\mathrm{min}}$, higher energy terms die off rapidly, and become insignificant, meaning we can reduce $N_{\mathrm{exp}}$ accordingly. In a perfect world, one would take each correlator to be fitted and trial different combinations of $t_{\mathrm{min}}$ and $N_{\mathrm{exp}}$, to find the point where adding further exponentials for a given $t_{\mathrm{min}}$ does not change the fit, leaving the extra fit parameters as their prior values. For a small fit to a handful of correlators, this method is a good one. For large fits to hundreds of correlators, varying $t_{\mathrm{min}}$ values for each rapidly becomes infeasible.

To this problem, we propose the following solution, allowing for a link between $t_{\mathrm{min}}$ and $N_{\mathrm{exp}}$ which is tailored to each correlator automatically. For each two-point correlator, we have an effective mass, discussed in Section~\ref{sec:example}. The uncertainty on this effective mass (the actual calculated effective mass, not the variable uncertainty we choose for our prior) gives a good order of magnitude estimate for the final uncertainty on our ground state energy in the fit. We also know that excited states are spaced above the ground state by roughly $\Lambda_{\mathrm{QCD}}$, and also that their amplitudes are likely of the same order of magnitude as the ground state amplitude, or smaller. So we can say that, for a given choice of $N_{\mathrm{exp}}$, the $t_{\mathrm{min}}$ value we should choose is given by the equation,
\begin{equation}
  e^{-\big(M_{\mathrm{eff}}+(N_{\mathrm{exp}}-1)\Lambda_{\mathrm{QCD}}\big)t_{\mathrm{min}}}\leq A\times \mathrm{err}(e^{-M_{\mathrm{eff}}t_{\mathrm{min}}}),
\end{equation}
where we can choose $A<1$ (e.g. $A=0.5$) to make our choice more conservative. What we are saying here is that, for a given $N_{\mathrm{exp}}$, we choose $t_{\mathrm{min}}$ such that, assuming the excited state amplitude is the same as the ground state, and they are spaced by $\Lambda_{\mathrm{QCD}}$, the highest excited state contribution at $t_{\mathrm{min}}$ is less than $A$ times the expected uncertainty on the ground state, and so irrelevant. This is in general a conservative estimate (i.e. giving a larger $t_{\mathrm{min}}$ than strictly required) for a number of reasons. Firstly, as mentioned, it assumes that the amplitudes for excited states are the same as the ground state. For a well chosen operator we would expect good ground state overlap, so the excited states should tend to be smaller. Secondly, it only looks at the relative size of the excited state contribution compared to the ground state uncertainty at $t_{\mathrm{min}}$. Since the excited state decays away must faster than the ground state, for $t>t_{\mathrm{min}}$ this will be even smaller, so a fit where $t_{\mathrm{min}}$ is 1 less than this recommended value would still in principle go through all data points, except with some tension on the first point.
By the same argument, we can take one fewer exponential for oscillating excited states in our fit, as we expect them to be roughly $\Lambda_{\mathrm{QCD}}$ above the non-oscillating ones. 

Once we have $t_{\mathrm{min}}$ for the two-point correlators, we can also take a maximal $t$ value, $t_{\mathrm{max}}$, to which the fit is much less sensitive in general. Methods for choosing $t_{\mathrm{max}}$ will be discussed below, but for now let's assume we have a $t_{\mathrm{min}}$ and $t_{\mathrm{max}}$ for each two-point correlator. We can derive the $t$ limits of the three-point correlators directly from this, removing potentially hundreds more variable choices from our fit.

From Eq.~\eqref{eq:3ptcorr}, we see that the three-point looks like $C_2^{M_1}(t)C_2^{M_2}(T-t)$, so we can simply say that we shall only take three-point values for $t$ for which $t^{M_1}_{\mathrm{min}}\leq t\leq t^{M_1}_{\mathrm{max}}$ and $t^{M_2}_{\mathrm{min}}\leq T-t\leq t^{M_2}_{\mathrm{max}}$. This could be considered a relatively conservative move, as higher order contributions to $C_3$ are often suppressed by small values of $J_{ij}^{kl}$. A less conservative alternative would be to calculate the two-point $t_{\mathrm{min}}$ values for $N_{\mathrm{exp}}+1$, and use \textit{those} values to derive the three-point $t$ range.

The entire purpose of this discussion is to reduce the variable parameters in our fit to a manageable number. Instead of trialling different $t_{\mathrm{min}}$ and/or $N_{\mathrm{exp}}$ for potentially hundreds of two, and three-point correlators, we have reduced the options, in a logical manner, down to a choice of $N_{\mathrm{exp}}$, and a couple of options to choose: $A$ and whether or not we choose $N_{\mathrm{exp}}+1$ for the three-point $t$ range. 
\subsection{Choosing $t_{\mathrm{max}}$}
As noted above, $t_{\mathrm{max}}$ is a much less important parameter than $t_{\mathrm{min}}$. In general, there is no reason to cut any data off the end of a correlator. This data will often be extremely noisy, so it won't contribute to the fit, but it won't do any harm either. However, as discussed in Sec.~\ref{sec:svd}, anything we can do to reduce the number of data points is a good idea, and throwing away noisy data which doesn't contribute to the fit is a good way to do this. 
\begin{figure}
  \includegraphics[width=0.44\textwidth]{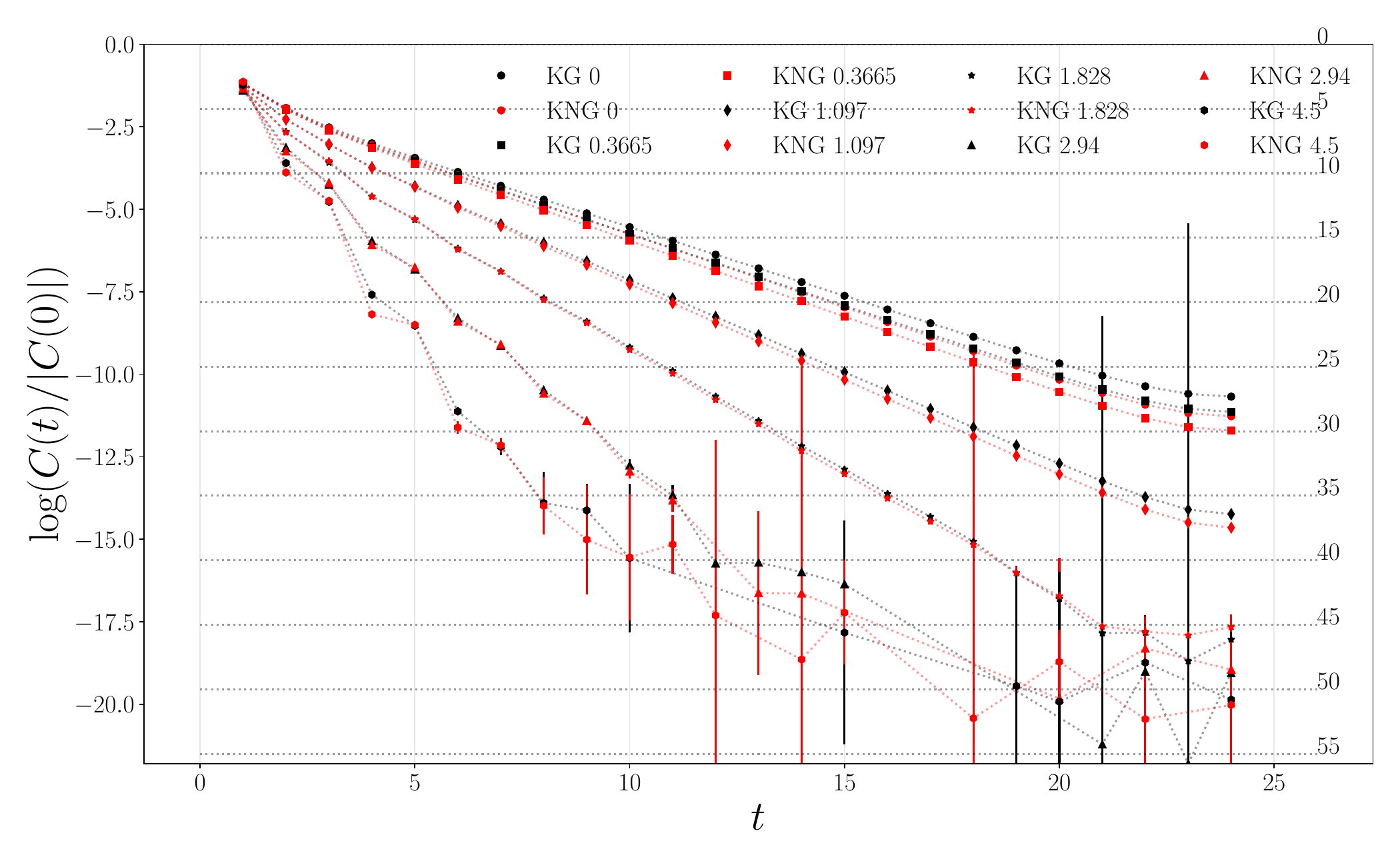}
  \caption{The log of $K$ correlators, for increasing momenta (twists). The black points show Goldstone (`G') spin-taste, and the red non-Goldstone (`NG').}
  \label{fig:log_C_vc}
\end{figure}

Figure~\ref{fig:log_C_vc} shows an example of a plot of $\log(C_2(t)/|C_2(0)|)$ for a number of different twist kaons. For the smaller twists, the correlator looks very clean, but for the larger twists, the correlator becomes very noisy at large $t$ values. A good way to chose $t_{\mathrm{max}}$ is to simply go through all the correlators by eye and pick a $t_{\mathrm{max}}$ for each where the data becomes visibly noisy. As with the $t_{\mathrm{min}}$ values discussed previously, this is fine for a few correlators, but it rapidly becomes infeasible to choose and vary so many individual parameters. We propose the solutions below, which could also be used to choose $t_{\mathrm{min}}$ values, as an alternative to the method discussed in Sec.~\ref{sec:tmin}.

Firstly, we note that, when we plot the log of the two-point correlator relative to itself at the origin, $\log(C_2(t)/|C_2(0)|)$, visible noise sets in around a similar value for a given correlator, regardless of heavy mass or twist. In the case of Figure~\ref{fig:log_C_vc}, we can see that all data below $\log(C_2(t)/|C_2(0)|)\approx-15$ is visibly noisy. For the cleaner correlators, their values never drop this low.

Similarly, the point where excited states are visibly decayed (i.e. the log plot becomes linear) also occurs near a common value for $\log(C_2(t)/|C_2(0)|)$. Using this information, we take tunable integer parameters $\tau^H_{\mathrm{min}}$ and $\tau^H_{\mathrm{max}}$, and calculate the corresponding $t_{\mathrm{min}}$ and $t_{\mathrm{max}}$ from points where
\begin{equation}
  e^{-a\lambda^{H/K} \tau^{H/K}_{\mathrm{min}}}<C(t)/|C(0)|<e^{-a\lambda^{H/K}\tau^{H/K}_{\mathrm{max}}},
\end{equation}
for each individual correlator. Here we take a scale, $\lambda$ with units of GeV (lattice spacing $a$ has units of $\mathrm{GeV}^{-1}$), simply to scale $\tau$ so that an integer change in $\tau$ has a sensible change in $C(t)$. For example, in Figure~\ref{fig:log_C_vc}, the $K$ meson have a mass of approximately $500$ MeV. So we take $\lambda^K=500$ MeV, meaning that a change of $\pm 1$ in $\tau^K$ corresponds to roughly 1 unit of $t$ in the twist 0 correlator, but less than this in higher twist (and so higher energy) correlators. For a $H$ meson, where the lowest $m_h=m_c$, we might choose $\lambda^{H}=2$ GeV.

The horizontal dotted lines and labels refer to the $\tau^K$ values that correspond to that $\log(C_2(t)/|C_2(0)|)$. As is perhaps obvious, the point of this method is that a change of 5 in the value of $\tau^K$ moves $t_{\mathrm{min/max}}$ by many more units for the lighter (low momentum) $K$s than for their heavier counterparts. We have created a way to adjust $t_{\mathrm{max}}$ across all $K$ correlators at once, taking into account their respective energies.   

As with the rest of this work, the point here is not to be able to choose the perfect $t_{\mathrm{max}}$ for each correlator every time. That can only be achieved through pricking them individually. But this hopefully provides a more nuanced suggestion for varying large numbers of $t_{\mathrm{max}}$ for different correlators than simply shifting $t_{\mathrm{max}}\to t_{\mathrm{max}}\pm 1$ across all correlators at once. 

\subsection{Binning over $t$ (coarse graining)}
Another option for reducing the number of data points in the fit is to bin correlators over $t$, or `coarse grain', them. This is a feature which is built into \textit{corrfitter}~\cite{corrfitter} (see documentation for implementation). It averages both the models and data over sets of $N_{\mathrm{cgr}}$ adjacent points in $t$. This is potentially a very powerful technique. Taking $N_{\mathrm{cgr}}=2$ on all data, for example, halves the number of data points, and so reduces the size of the covariance matrix by a factor of four\footnote{Or two if you consider its symmetry.}. The cost is of course an effective loss of data, and so an increase in uncertainty.

It is often not worth the loss in precision to coarse grain all data, but it can be worth taking a more targeted approach. For example, when fitting two, and three-points simultaneously, one might not coarse grain the two points, and choose one $T$ value for the three-points which is also left unaffected, then coarse grain the remaining three-points, taking $N_{\mathrm{cgr}}=2$. Given the method for choosing $t_{\mathrm{min}}$ and $t_{\mathrm{max}}$ above, one may find that some three-point functions only contain 2 or 3 $t$ values. It may be valuable to set a condition where coarse graining is only activated on correlators longer than a certain threshold. 

\section{Conclusions}\label{sec:conclusions}
This work is intended as a source of helpful suggestions for those seeking to perform very large, correlated fits to two, and three-point correlation functions, chiefly, but not necessarily, using \textit{gvar}, \textit{lsqfit} and \textit{corrfitter}~\cite{lsqfit,gvar,corrfitter}. It is by no means comprehensive, but reflects some personal observations and (hopefully) useful tips. There are many other approaches that may be useful and are not discussed here, such as chaining and marginalisation. We once again direct the reader to the \textit{lsqfit} and \textit{corrfitter} documentation for more information. 

\section{Acknowledgements}
We are grateful to the MILC collaboration for the use of  their  configurations  and  their  code, which we use to generate quark propagators and construct correlators. We would also like to thank G. P. Lepage, R. Lewis and J. Harrison for useful discussions. Computing was done on the Cambridge Service for Data Driven Discovery (CSD3) supercomputer, part of which is operated by the University of Cambridge Research Computing Service on behalf of the UK Science and Technology Facilities Council (STFC) DiRAC HPC Facility. The DiRAC component of CSD3 was funded by BEIS via STFC capital grants and is operated by STFC operations grants. We are grateful to the CSD3 support staff for assistance. Funding for this work came from the Natural Sciences and Engineering Research Council of Canada.

\bibliography{fitting}
\end{document}